\documentclass[twocolumn]{aastex63}
\usepackage[T1]{fontenc}
\usepackage{color,epsfig,graphics,graphicx,rotating,comment,amssymb,stackengine}
\usepackage{natbib,lipsum,atbegshi,lineno}
\usepackage{booktabs}
\usepackage{sidecap}
\usepackage{ulem}

\definecolor{brown}{rgb}{0.6,0.4,0.2}
\definecolor{purple}{rgb}{0.5,0,0.5}
\shortauthors{Hwangbo et al.}  

\def\msun{M$_{\odot}$}
\def\one{{\,\sc i}}
\def\two{{\,\sc ii}}

\newcommand{\kms}{km\,s$^{-1}$}

\newcommand{\rbc }{SN\,2024rbc}

\newcommand\new[1]{}

\begin{document}

\title{Near-Infrared and Optical Observations of \rbc: The First Early Detection of CO and Dust in a Type Ib Supernova}

\correspondingauthor{Ryan Hwangbo}
\email{ryanhwb@berkeley.edu}

\author[0000-0002-9789-0649]{Ryan Hwangbo}\affil{Department of Physics, University of California, Berkeley, CA 94720-3411, USA}\affil{SETI Institute, 339 Bernardo Ave., Ste. 200, Mountain View, CA 94043, USA}
\author[0000-0001-7488-4337]{Jeonghee Rho}\affil{SETI Institute, 339 Bernardo Ave., Ste. 200, Mountain View, CA 94043, USA}
\author[0000-0002-7352-7845]{Aravind P. Ravi}\affil{Department of Physics and Astronomy, University of California, 1 Shields Avenue, Davis, CA 95616-5270, USA}
\author[0000-0001-7488-4337]{Seong Hyun Park}\affiliation{Department of Physics and Astronomy, Seoul National University, Gwanak-ro 1, Gwanak-gu, Seoul, 08826, South Korea}
\author[0000-0003-2946-9390]{Harim Jin}\affiliation{Argelander-Institut f\"ur Astronomie, Universit\"at Bonn, Auf dem H\"ugel 71, 53121 Bonn, Germany}\affiliation{Max Planck Institute for Astrophysics, Karl-Schwarzschild-Straße 1, 85748 Garching bei München, Germany}
\author[0000-0001-7488-4337]{Sung-Chul Yoon}\affiliation{Department of Physics and Astronomy, Seoul National University, Gwanak-ro 1, Gwanak-gu, Seoul, 08826, South Korea}
\author[0000-0003-2824-3875]{T. R. Geballe}\affiliation{Gemini Observatory/NSF's National Optical-Infrared Astronomy Research Laboratory, 670 N. Aohoku Place, Hilo, HI 96720, USA}
\author[0000-0002-2445-5275]{Ryan Foley}\affil{Department of Astronomy and Astrophysics, University of California, Santa Cruz, CA 95064, USA}
\author[0000-0002-5748-4558]{Kirsty Taggart}\affil{Department of Astronomy and Astrophysics, University of California, Santa Cruz, CA 95064, USA}
\author[0000-0002-5680-4660]{Kyle W. Davis}\affil{Department of Astronomy and Astrophysics, University of California, Santa Cruz, CA 95064, USA}
\author[0000-0002-1092-6806]{Kishore C. Patra}\affil{Department of Astronomy and Astrophysics, University of California, Santa Cruz, CA 95064, USA}
\author[0000-0002-1481-4676]{S. Tinyanont}\affil{Department of Astronomy and Astrophysics, University of California, Santa Cruz, CA 95064, USA}\affil{National Astronomical Research Institute of Thailand, 260 Moo 4, Donkaew, Maerim, Chiang Mai, 50180, Thailand}
\author[0000-0003-1546-6615]{Jesper Sollerman}\affil{Department of Astronomy, Stockholm University, 10691 Stockholm, Sweden}
\author[0000-0001-6797-1889]{Steve Schulze}\affil{Center for Interdisciplinary Exploration and Research in Astrophysics (CIERA), Northwestern University, 1800 Sherman Ave, Evanston, IL 60201, USA}
\author[0000-0002-2249-0595]{Natalie LeBaron}
\affil{Department of Astronomy, University of California, Berkeley, CA 94720-3411, USA}
\affil{Berkeley Center for Multi-messenger Research on Astrophysical Transients and Outreach (Multi-RAPTOR), University of California, Berkeley, CA 94720-3411, USA}
\author[0000-0002-7866-4531]{Chang Liu}\affil{Center for Interdisciplinary Exploration and Research in Astrophysics (CIERA), Northwestern University, 1800 Sherman Ave, Evanston, IL 60201, USA}\affil{Department of Physics and Astronomy, Northwestern University, 2145 Sheridan Rd, Evanston, IL 60208, USA}
\author[0000-0002-5740-7747]{Charles D. Kilpatrick}\affiliation{Center for Interdisciplinary Exploration and Research in Astrophysics (CIERA), Northwestern University, 1800 Sherman Ave, Evanston, IL 60201, USA}
\author[0000-0002-6230-0151]{David O. Jones}\affil{Institute for Astronomy, University of Hawai$'$i, 640 N Aohoku Place, Hilo, HI 96720, USA}
\author[0009-0006-5214-0736]{C. Tanner Murphey}\affil{Department of Astronomy, University of Illinois, Urbana, IL 61801, USA}\affil{Center for Astrophysical Surveys, National Center for Supercomputing Applications, Urbana, IL 61801, USA}\affil{Illinois Center for Advanced Studies of the Universe, Urbana, IL 61801, USA}

\begin{abstract} 
\noindent 
We present optical and near-infrared (NIR) observations of the Type Ib supernova (SN) 2024rbc. Emission from the first CO overtone, resting on a dust continuum at $2.3-2.4$ $\mu$m, was observed at 62 days post-explosion. 
\new{The CO band heads are not resolved, and the emission appears broad, lacking sharp spectral features.}
This is the first observation of CO in the ejecta of a Type Ib SN reported in literature. The spectra of \rbc\ exhibit strong He I lines and numerous neutral and ionized metal lines. Comparing the spectral evolution of \rbc\ to other Type Ib, Ic, and IIb SNe indicates it is a Type Ib SN. We compare the velocities of key optical lines to examine the evolution of the ejecta. Additionally, fitting SN light curve models of helium star progenitors computed with the STELLA code to photometric observations indicates a $^{56}$Ni mass of $0.07$ $M_\odot$ and an ejecta mass of $1.7$ $M_\odot$.
Fitting a LTE model to the CO overtone \new{implies a CO} mass of 5.2 $\times$ 10$^{-4}$ $M_{\odot}$, a \new{CO} temperature of $4040$ K, and a \new{CO} velocity width of $5905$ \kms. We also fitted a modified blackbody model to the dust continuum, deriving a dust temperature of $910$ K and a \new{dust} mass of $1.3$ $\times$ $10^{-3}$ $M_{\odot}$. 
\new{The CO provides direct evidence for the onset of dust formation, and the observed dust continuum likely originates from newly formed dust in the ejecta. However, robust dust mass estimates require MIR observations.}

\vspace{5ex}

\end{abstract}

\section{Introduction}\label{Section: Introduction}
In the present Universe, asymptotic giant branch (AGB) stars are the primary dust producers. However, the high dust content ($>10^8$ $M_\odot/\text{galaxy}$) of high-redshift ($z\geq6$) galaxies \citep{bertoldi03, robson04, beelen06} is not explained by AGB stars. The total dust formation ($\sim$$10^{-3}$ $M_{\odot}/\text{star}$, see \citealt{michaowski15} for more details) and maturation time of AGB stars ($10^8-10^{10}$ years) are in conflict with the observed quantity of dust and the age of the Universe at such high redshifts ($<10^9$ years). 

A more plausible source of large dust masses in the early Universe \new{is} massive Population II/III stars. Their short stellar lifetimes, ending in core-collapse supernovae (CCSNe) or pair-instability supernovae (PISNe), and high metal production are conducive to the rapid enrichment of the interstellar medium (ISM) with dust-forming elements \citep{todini01, nozawa03}. Observations of nearby \new{CCSNe} remnants have revealed that \new{they} are an important source of dust (Cas A, \citealt{CasADust_de_looze17} \& \citealt{rho08}; SN 1987A, \citealt{1987ADust_wesson15} \& \citealt{1987ADust_matsuura15}; Crab Nebula, \citealt{CrabDust_de_looze19}; SNR G54.1+0.3, \citealt{G54Dust_temim17} \& \citealt{G54Dust_rho18}). These observations, in conjunction with theoretical models of CCSNe, have shown formed dust masses ranging from $\sim$$10^{-2}$ to $1$ $M_\odot/\text{SN}$ \citep{todini01, nozawa03, bianchi07, gomez09, cherchneff10}. As such, continued investigation of CCSNe as a dust production mechanism is critical.

Type Ib and Ic SNe are two types of CCSNe \new{distinguished by their spectral features}. The former lacks hydrogen lines, while the latter lacks both hydrogen and helium lines \citep{filippenko97, matheson01, gal-yam17}. The absence of these lines indicates that the progenitor star had lost most, if not all, of the corresponding envelope layers during its evolution. These two types (along with Type IIb SNe) are collectively referred to as stripped-envelope supernovae (SESNe), which presents a more focused area of interest for SN-mediated dust formation. 

Key envelope-stripping mechanisms in the progenitors of SESNe include stellar winds and binary interactions \citep{yoon10, yoon15, yoon17, aguilera-dena22, sun22, hirschi25, jin26}. While stellar winds are generally only sufficient in high-mass stars (M$_{\text{ZAMS}}>25$ $M_{\odot}$), binary interactions are fairly mass-independent. Past decades of observational and theoretical study of SESNe have pointed to binary interactions being the primary pathway \citep{podsiadlowski92, wellstein99, eldridge08, yoon10, drout11, lyman16, yoon17, taddia18, sun22}.

Models of helium stars (He stars; evolved stars that have lost their hydrogen envelopes) with masses between $4-12$ $M_{\odot}$ built by \cite{dessart20} using the MESA code \citep{Paxton2011, Paxton2013, Paxton2015, Paxton2018} show that less massive progenitors retain helium-rich envelopes and reproduce the colors, line widths, and line strengths representative of Type Ib SNe. Conversely, more massive progenitors were found to lose most of their helium via stellar winds and generate spectra matching Type Ic SNe. \cite{yoon10} explored detailed binary evolution models for SESN progenitors, covering a wide range of initial masses for the primary components ($12-25$ $M_{\odot}$). Assuming solar metallicity, final He star masses were between $1.5$ and $7.1$ $M_{\odot}$, and low helium content ($<0.5$ $M_{\odot}$) was most likely at the extremes ($<2.0$ $M_{\odot}$ and $>5.5$ $M_{\odot}$). A thin hydrogen layer was found for a narrow range of final masses ($3-3.7$ $M_{\odot}$). This suggests that Type Ib and Type Ic SNe come from similar progenitors but differ in degree of envelope stripping due to factors like mass, metallicity, and dynamical history. Further examination of the properties of these SESNe and their progenitors is necessary to better understand the causes of this divergence.

Examining newly discovered SESNe for evidence of early dust formation is essential to testing the possibility of CCSN-mediated dust formation in the early Universe. As an indicator of molecular cooling and chemistry in the ejecta, CO is key to this search \citep{sarangi18, G54Dust_rho18, rho21}. The proliferation of ground-based spectrographs capable of rapidly targeting new discoveries for observation has opened the door to monitoring the first CO overtone ($>2.29$ $\mu$m). Observations of Type Ib and Ic SNe have shown rapid dust formation accompanied by detections of CO emission \citep{rho21, rho18sn, ravi23, liu92}. 

Among SESNe, Type Ib and Ic SNe have particular advantages in the search for early CO and dust formation as they dim faster than Type IIb SNe. This reduces the emission window for ionizing Compton electrons and thermal radiation that inhibit molecule formation, shortening the condensation time for CO and dust grains \citep{nozawa08}. Combined with the proportionately C- and O-rich ejecta of these SNe, it is plausible that significant CO and dust formation occurs shortly after the explosion. Indeed, observations of CO in Type Ic SNe have been reported at least as early as 63 days post-explosion (2020oi, \citealt{rho21}; 2016adj, \citealt{banerjee18}; 2021krf, \citealt{ravi23}). However, no such observation of CO has been reported in a Type Ib SN.

In this paper, we detail observations of Type Ib \rbc, which exhibited CO spectral features and warm dust emission as early as 62 days post-explosion.  We present the discovery of \rbc\ in Section \ref{Subsection: Discovery}, optical photometry in Section \ref{Subsection: Optical Photometry}, optical spectroscopy in Section \ref{Subsection: Optical Spectroscopy}, and NIR spectroscopy in Section \ref{Subsection: NIR Spectroscopy}.

We subsequently present our estimation of the explosion date of \rbc\ in Section \ref{Subsection: Explosion Date}, extinction correction in Section \ref{Subsection: Extinction}, light curves in Section \ref{Subsection: Light Curves}, bolometric luminosity estimation and progenitor fitting in Section \ref{Subsection: Bolometric Luminosity and Light Curve Fitting}, optical spectra in Section \ref{Subsection: Optical Spectra Analysis}, and NIR spectra in Section \ref{Subsection: NIR Spectra Analysis}. 

We then discuss our comparison of \rbc\ to other Type Ib and Ic supernovae. Light curves are compared in Section \ref{Subsection: Light Curve Comparison}, the optical spectra in Section \ref{Subsection: Optical Spectra Evolution and Comparison}, and the NIR spectra in Section \ref{Subsection: NIR Spectra Evolution and Comparison}. In Section \ref{Subsection: Line Velocity Analysis}, we analyze the velocity profiles and evolution of several key spectral lines. Section \ref{Subsection: CO and Dust Modeling} presents our CO models and constrains the mass, velocity, and temperature of the CO and dust in \rbc. Finally, a summary of the key takeaways from this paper is provided in Section \ref{Section: Conclusions}, and 
reductions of the Gemini GNIRS data using the \texttt{xdgnirs}, \texttt{pypeit}, and \texttt{Figaro} packages are compared in the Appendix.

\section{Observations}\label{Section: Observations}
\subsection{Discovery}\label{Subsection: Discovery}
\begin{figure}
\centering
\includegraphics[scale=0.2]{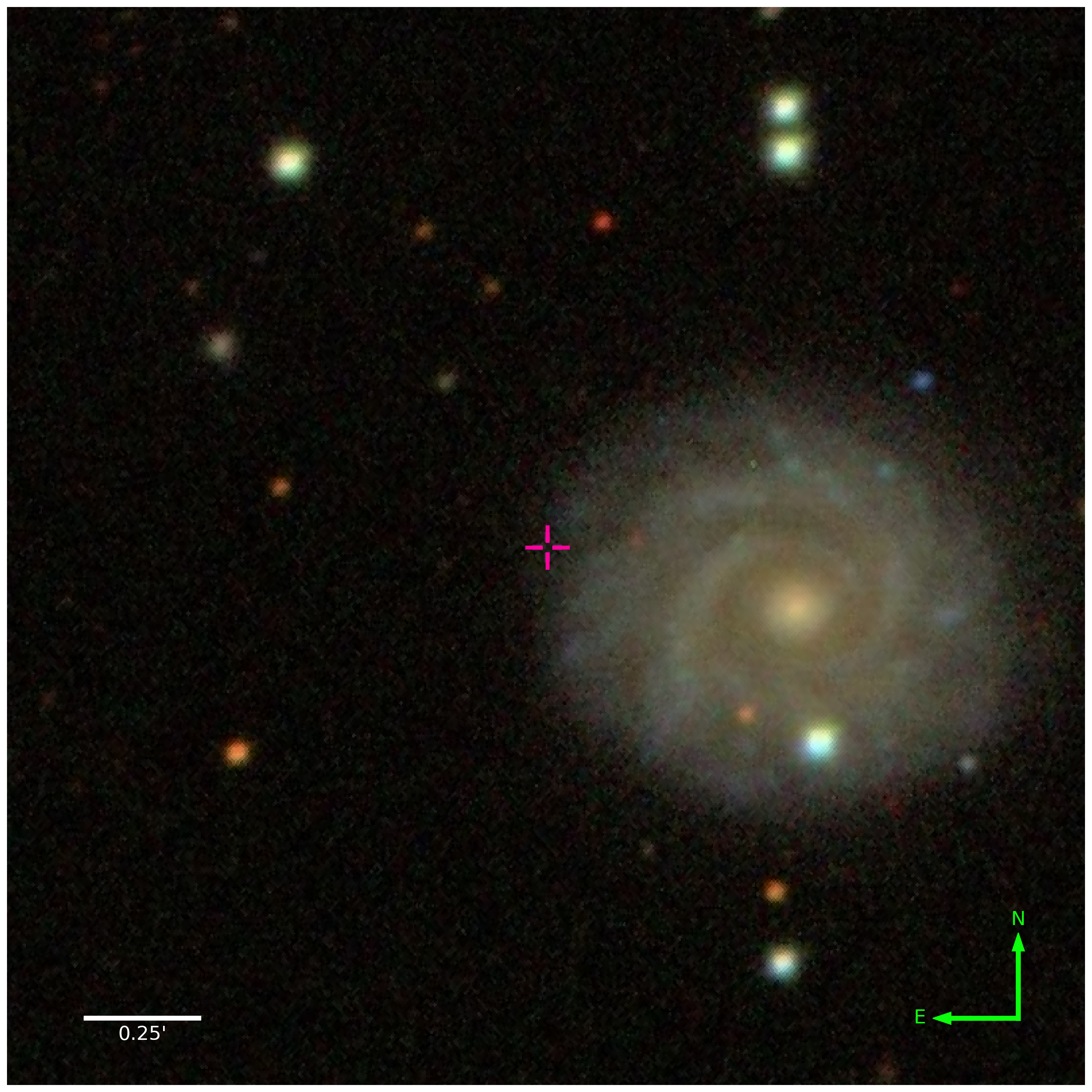}
\caption{\new{A 2.39$'$ $\times$ 2.39$'$ Sloan Digital Sky Survey (SDSS) Data Release 9 image (\emph{gri} composite) centered on the position of \rbc\ (purple cross), which was not visible at the time this image was taken. The spiral galaxy NGC 39 is also visible in the figure.}}
\label{Figure:SDSSImage}
\end{figure}

\begin{table}
    \begin{center}
    \caption{Properties of \rbc}\label{Table:KeyParameters}
    \vspace{-10pt}
        \begin{tabular}{l|cc}
        \hline \hline 
        Parameter &  &\\
        \hline
        R.A.\,(J2000)                                      &00$^{h}$12$^{m}$21$^{s}$.45&\\
        Dec.\,(J2000)                                      &+31$^{\circ}$00$'$48$''$.11&\\
        Distance (Mpc)                                     &66.80 $\pm$ 4.69&\\
        Redshift ($z$)                                     &0.016&\\
        E($B-V$) (MW, host; mag)                           &0.0619, $<0.02$\\
        E($B-V$) (total; mag)                              &0.0619 $\pm$ 0.0011&\\
        Explosion Date (MJD)                               &60525.15 $\pm$ 0.20 \\
        \hline
        pre-explosion He star mass  (\msun) & 3.1 \\
        {\bf Ejecta} Mass ($M_{\odot}$)                          & 1.7 &\\
        Explosion Energy (10$^{51}$ erg)  & 1.0 &\\
        Ni Mass ($M_{\odot}$)                              & 0.07 &\\
        Ni Mixing Fraction ($f_{\text{m}}$)                & 0.5 &\\
        Final Mass ($M_{\odot}$)                           & 3.1 &\\
        Progenitor Mass at ZAMS ($M_{\odot}$)              & $\sim$11 &\\
        CSM Mass ($M_{\odot}$)                             & $<1.8\times10^{-2}$\\
        \hline \hline 
        \end{tabular}
        \end{center}
\renewcommand{\baselinestretch}{0.8}
\vspace{-7pt}
\footnotesize{The progenitor star parameters are indicative, not definitive. See Section \ref{Subsection: Bolometric Luminosity and Light Curve Fitting} for more details.}
\end{table}

{\bf \rbc} (ZTF24aaymkrs) was discovered by the Zwicky Transient Facility (ZTF) on 2024 August 3 at 08:20:55 UTC \citep{2024TNSTR2747....1D} using the ZTF Camera \citep{ZTFC}. 
We identify NGC 39 as the host galaxy. 
The NASA/IPAC Extragalactic Database\footnote{\url{https://ned.ipac.caltech.edu/}} (NED) lists the spiral galaxy NGC 39 at a redshift of $z = 0.0162$ \citep{huchra99} and at a Hubble distance of $66.80 \pm 4.69$ Mpc \citep[see also][]{haynes18}. 
We adopt this distance for \rbc\ as well. \new{With respect to NGC 39, \rbc\ is north by $2.6 \pm 0.2$ kpc ($\sim$$8.2^{\prime\prime}$) and east by $12.5 \pm 0.9$ kpc ($\sim$$38.9^{\prime\prime}$), as shown in Figure \ref{Figure:SDSSImage}.}
\rbc\ was classified as a Type Ib supernova ($z=0.016$) using an optical spectrum obtained by the Spectral Energy Distribution Machine \citep[SEDM;][]{SEDM18} on 2024 August 13 at 08:56:16 UTC \citep{2024TNSCR2931....1S}.
These properties are listed in \autoref{Table:KeyParameters}.

\subsection{Optical Photometry}\label{Subsection: Optical Photometry}

\begin{figure*}
\centering
\includegraphics[scale=0.36]{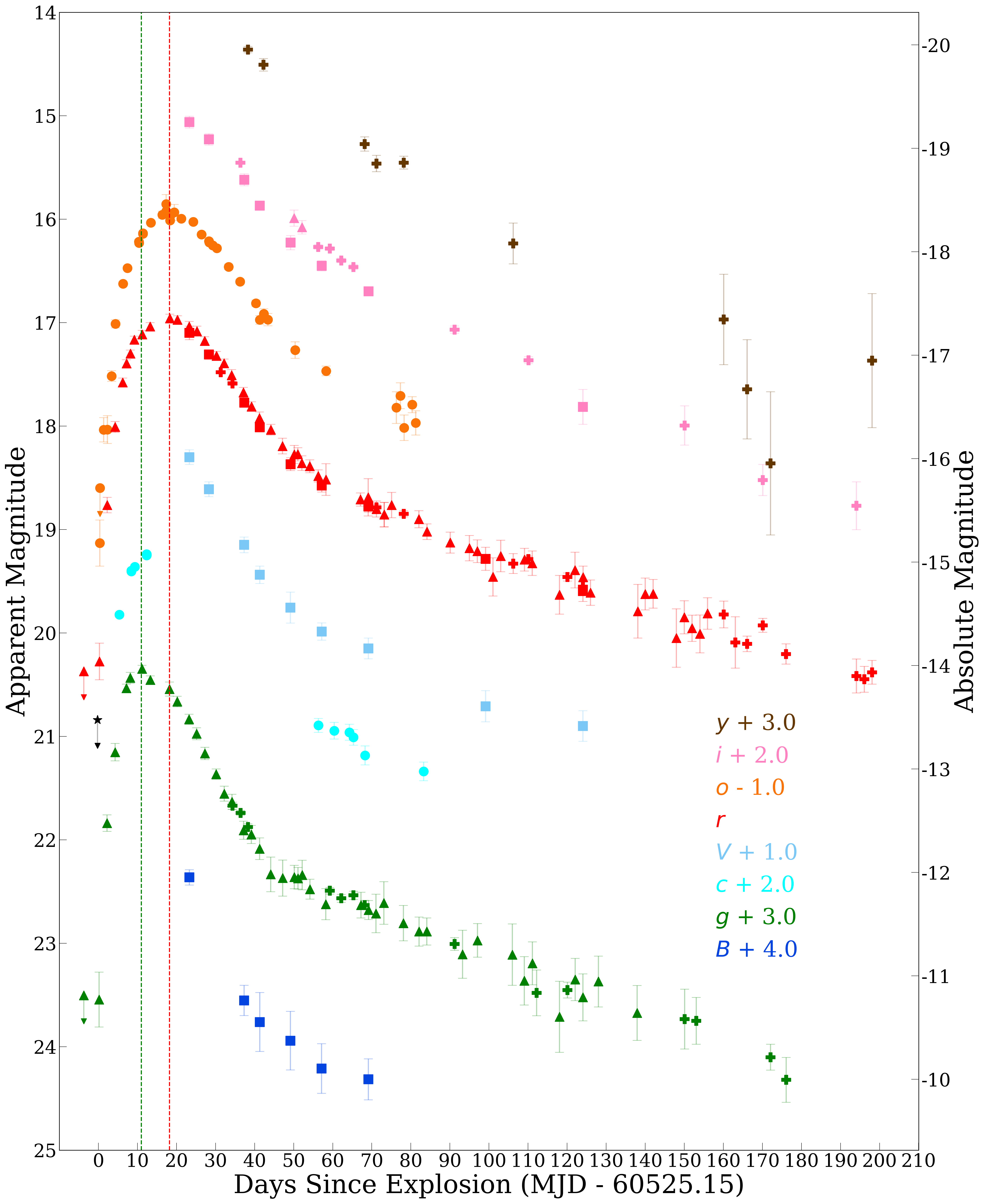}
\caption{
Light curves of \rbc\ in the \emph{BgcVroiy} bands (ordered by \new{wavelength}). ZTF data are in triangles, ATLAS in circles, Nickel in squares\new{, and Pan-STARRS in "+" markers}. The \emph{c} and \emph{o} bands correspond to the cyan and orange ATLAS filters. The apparent magnitudes are in AB, and the absolute magnitudes are shown on the right. The vertical dashed lines indicate the peaks for the $g$ and $r$ bands. \new{The last non-detection upper limits in the unfiltered (black, offset $=0$) and $gro$ bands are marked using downward arrows.} 
}
\label{Figure:Photometry}
\end{figure*}

Our optical photometry of \rbc\ \new{extends} from $<$$1$ to $\sim$$200$ days post-explosion. \autoref{Figure:Photometry} displays these light curves. The photometry is most complete in the $g$ and $r$ bands.

The ZTF Camera, mounted on the 48 inch Samuel–Oschin telescope (Schmidt type) at the Palomar Observatory, collected 53, 2, and 60 epochs in the $g$, $i$, and $r$ bands, respectively.

Additionally, the Asteroid Terrestrial-impact Last Alert System (ATLAS) 0.5 meter Wright–Schmidt telescope at Haleakalā, Maui (ATLAS–HKO) also observed \rbc. In total, 13 epochs in the cyan ATLAS band and 49 epochs in the orange ATLAS band were obtained with ATLAS–HKO.

Furthermore, eight epochs in each of the $B$ and $i$ bands and nine epochs in each of the $V$ and $r$ bands were obtained with the 1 meter Nickel telescope at the Lick Observatory. The images were calibrated using bias and sky flat-field frames following standard procedures. Point-spread function (PSF) photometry was performed and calibrated relative to the Panoramic Survey Telescope and Rapid Response System \citep[Pan-STARRS;][]{Flewelling20}.

\new{Lastly, through a small data collaboration with the Young Supernova Experiment \citep[YSE;][]{YSE21, PanSTARRS23}, we obtained 18, 24, 14, and 17 epochs of observation by Pan-STARRS at the Haleakala Observatory in the g, r, i, and y bands, respectively. The pertinent reduction process is detailed in Section 2.4 of \cite{PanSTARRS23}.}

\subsection{Optical Spectroscopy}\label{Subsection: Optical Spectroscopy}

\begin{table}
\begin{center}
\caption{Optical and NIR Spectroscopy of \rbc}
\label{Table:Spectroscopy}
\vspace{-0.05cm}
\resizebox{\columnwidth}{!}{
\hspace{-2.1cm}
\begin{tabular}{ccccccc}
\hline 
\hline
         Date        &       MJD      &      Epoch$^{a}$       &       Telescope       &     Instrument   \\
\hline
          2024-08-09 &          60531 &           6      &          Keck I       &          LRIS    \\
          2024-08-13 &          60535 &          10      &          ZTF P60      &          SEDM    \\
          2024-08-15 &          60537 &          12      &          NOT          &          ALFOSC  \\
          2024-08-27 &          60549 &          24      &          Lick         &          Kast    \\
          2024-09-08 &          60561 &          36      &          Lick         &          Kast    \\
 \textbf{2024-09-12}$^b$ & \textbf{60565} & \textbf{40}      & \textbf{Keck II}      & \textbf{NIRES}   \\
          2024-09-13 &          60566 &          41      &          Lick         &          Kast    \\
 \textbf{2024-09-19}$^b$ & \textbf{60572} & \textbf{47}      & \textbf{IRTF}         & \textbf{SpeX}    \\
          2024-10-03 &          60586 &          61      &          Lick         &          Kast    \\
 \textbf{2024-10-04}$^b$ & \textbf{60587} & \textbf{62}      & \textbf{Gemini-N}     & \textbf{GNIRS}   \\
\hline
\hline
\end{tabular}}
\end{center}
\vspace{-0.25cm}
\footnotesize{$^a$ The epoch is the measured from the explosion date ($\text{MJD}- 60525.15$), which is described further in Section \ref{Subsection: Explosion Date}.}\\
\footnotesize{$^b$ Rows corresponding to NIR observations are marked in bold.}
\end{table}

\begin{figure*}
\centering
\includegraphics[scale=0.375]{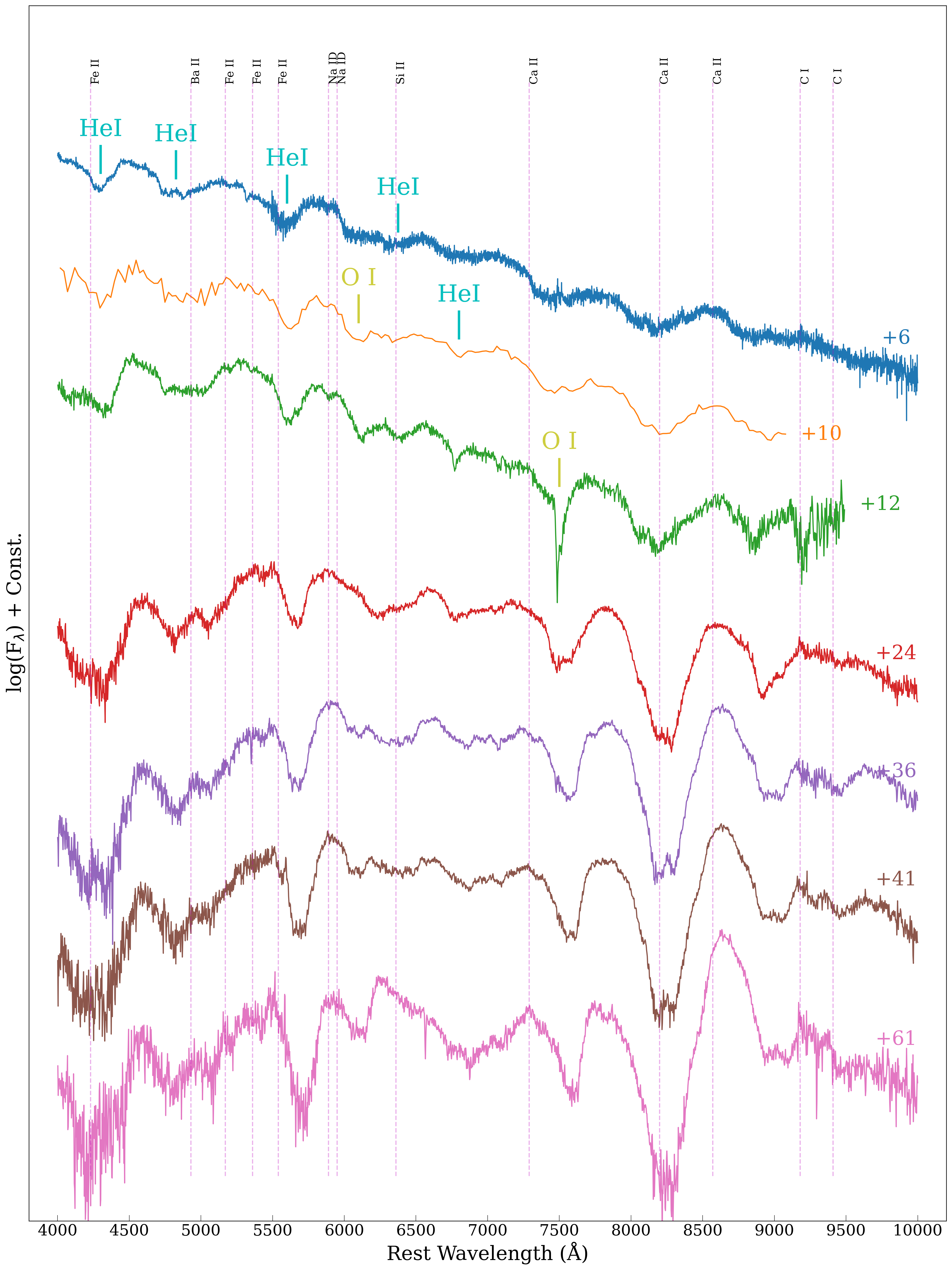} 
\caption{The optical spectra of \rbc. The lines of important metals are marked at rest wavelength (in standard temperature and pressure air) in magenta. The neutral helium and oxygen features are marked at the approximate absorption minima in cyan and yellow, respectively. The ionized calcium line at $\sim$$8567$ \AA\ refers to the mean of the NIR Ca II triplet (8498, 8542, and 8662 \AA).}
\label{Figure:OpticalSpectra}
\end{figure*}

We observed the optical spectrum of \rbc\ at seven epochs using four different instruments. \autoref{Table:Spectroscopy} summarizes these observations. The optical spectra
are shown in \autoref{Figure:OpticalSpectra}.
All spectra in this work were extracted from 2D frames using the optimal extraction algorithm \citep{horne1986}.

On 2024 August 9, we obtained the first optical spectrum of \rbc\ using the Low Resolution Imaging Spectrometer \citep[LRIS;][]{LRIS95} on the Keck I 10 meter telescope at the W. M. Keck Observatory. Observations were conducted with a 1\farcs0-wide slit, using the 400/3400 grism for the blue side and the 400/8500 grating for the red side. The data were reduced using \texttt{pypeit} \citep{pypeit:joss_pub}, a semi-automatic spectroscopic data reduction pipeline. The code for \texttt{pypeit} can be found on Zenodo \citep{pypeit:zenodo}.

The very low-resolution (R$\,\sim\,$100) integral field unit (IFU) spectrograph SEDM on the Palomar 60 inch telescope (P60) at the Palomar Observatory observed \rbc\ on 2024 August 13. The reduced spectrum \citep{rigault19} was retrieved from the Transient Name Server (TNS).

Following this, the Alhambra Faint Object Spectrograph and Camera\footnote{\url{https://www.not.iac.es/instruments/alfosc}} (ALFOSC) on the Nordic Optical Telescope (NOT) at the Roque de los Muchachos Observatory observed \rbc\ on 2024 August 15. The observation used Grism 4 with a 1\farcs3-wide slit and consisted of a 1650 second exposure, during which the slit was aligned along the parallactic angle. This spectrum was also reduced using \texttt{pypeit}.

Finally, optical spectra were obtained with the Kast Double Spectrograph \citep{miller1993lick} on the Shane 3 meter telescope at the Lick Observatory on 2024 August 27, September 8, September 13, and October 3. We used the {\tt UCSC Spectral Pipeline}\footnote{\url{https://github.com/msiebert1/UCSC\_spectral\_pipeline}}
\citep{siebert20}, a custom data-reduction pipeline based on procedures outlined by \citet{foley03} and \citet[][and references therein]{silverman2012}. The two-dimensional spectra were bias-corrected, flat-fielded, adjusted for varying gains across different chips and amplifiers, and trimmed. Wavelength calibration was performed using internal comparison-lamp spectra, with linear shifts applied by cross-correlating observed night-sky lines in each spectrum to a master night-sky spectrum. Flux calibration and telluric correction were performed using standard stars observed at a similar airmass to the science exposures. 

\subsection{Near-Infrared Spectroscopy}\label{Subsection: NIR Spectroscopy}

\begin{figure*}
\centering
\includegraphics[scale=0.375]{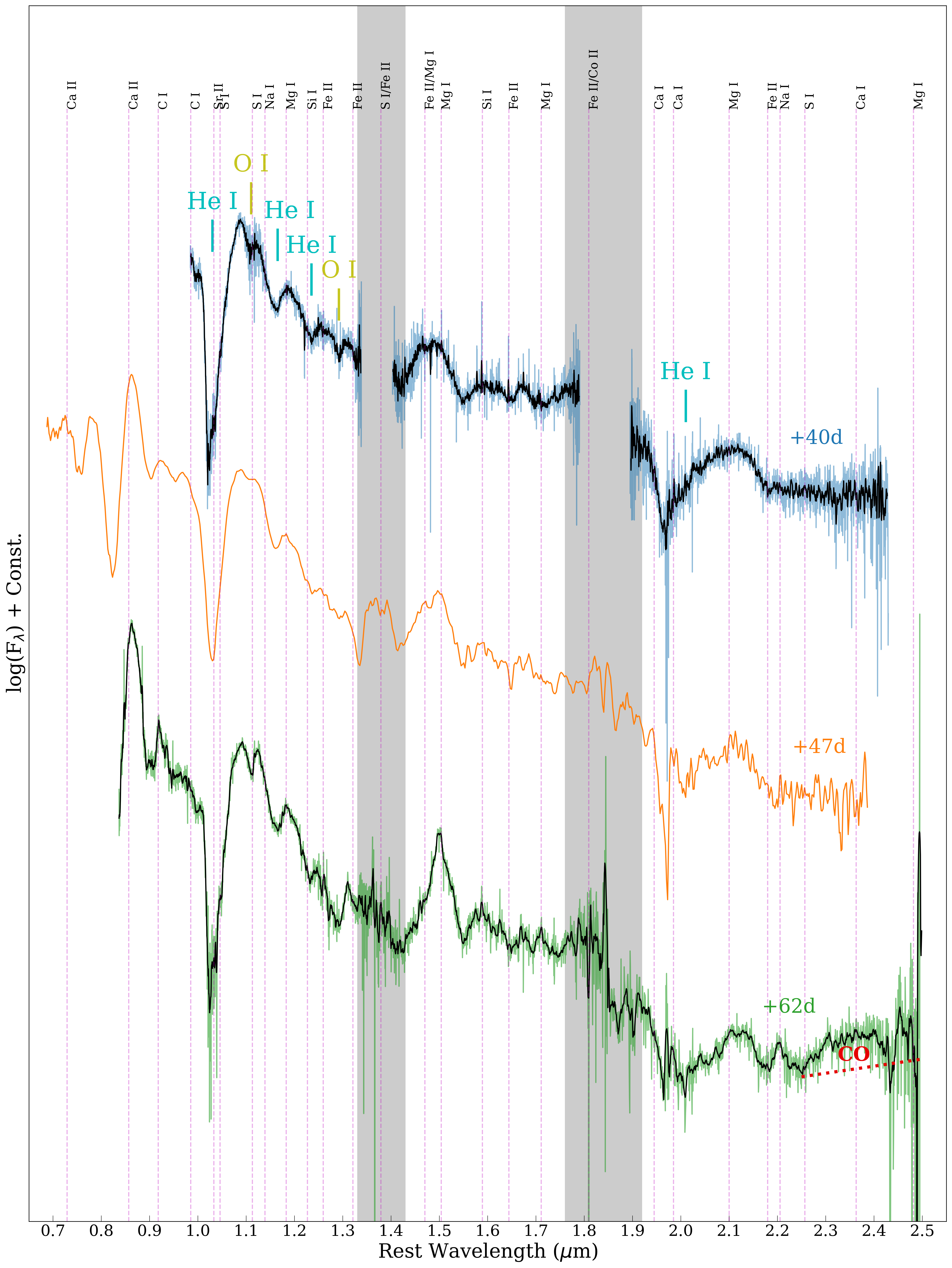}
\caption{The NIR spectra of \rbc. Key spectral lines are indicated at the top; lines that overlap on the scale of this figure are marked together. The lines of important metals are marked at rest wavelength (in vacuum) in magenta. The neutral helium and oxygen features are marked at the approximate absorption minima in cyan and yellow, respectively. Smoothed $+40$d and $+62$d spectra (black) are overlaid on the original spectra to improve feature visibility. At $+62$d, an estimate of the dust continuum is indicated by a red dotted line, and the velocity-broadened first CO overtone is marked above. The gray regions indicate unreliability due to low atmospheric transmission.}
\label{Figure:NIRSpectra}
\end{figure*}

\autoref{Figure:NIRSpectra} shows the NIR spectra of \rbc\ we collected. \autoref{Table:Spectroscopy} summarizes these observations as well. 

The earliest spectrum was obtained on 2024 September 12 with the Near-Infrared Echellette Spectrometer \citep[NIRES;][]{NIRES_wilson04} on the Keck II 10 meter telescope at the W. M. Keck Observatory. NIRES provides a resolving power of R$\,\sim\,$2700 over a wavelength range of $0.8-2.4$ $\mu$m, divided into six orders. The \texttt{Spextool} package \citep{2004PASP..116..362C} was used to reduce the NIRES data. To improve the visibility of key spectral features, the spectrum was smoothed using a third-order Savitzky–Golay filter with a 19-pixel window.

\rbc\ was subsequently observed on 2024 September 19 with the SpeX spectrograph \citep{SPEX} at the NASA Infrared Telescope Facility (IRTF) on Mauna Kea (PI: A.~P. Ravi). SpeX observed the wavelength range $0.7-2.52$ $\mu$m in its low-resolution prism mode (R$\,\sim\,$82) with a $0\farcs8 \times 15$\arcsec\ slit. The data were reduced using \texttt{Spextool}; telluric correction and flux calibration were applied using subroutines. We smoothed the spectrum using a 2-pixel full width at half maximum (FWHM) Gaussian.

The final NIR spectrum was obtained on 2024 October 4 with the Gemini Near-Infrared Spectrograph \citep[GNIRS;][]{elias06d, elias06p} on the 8.1 meter Frederick C. Gillett Gemini–North telescope at the Gemini Observatory on Mauna Kea. These data were collected as part of our Gemini program (GN-2024B-Q-216, PI: S.~H. Park). GNIRS was configured to its 32 lines~mm$^{-1}$ short cross-dispersed (SXD) mode with a 0\farcs45 slit. \rbc\ was observed in the standard stare/nod-along-slit mode with a nod angle of 3\farcs0. The instrument achieved a spectral resolution of R$\,\sim\,$1200 with a total integration time of \new{3600} seconds. 

The GNIRS dataset was reduced using three methods: (1) \texttt{xdgnirs}, a PyRAF-based data-reduction pipeline \citep[see][]{mason15, ravi23, park25}; (2) \texttt{Figaro} \citep{shortridge92}, combined with standard IRAF \citep{tody86} tools for manual order-by-order data reduction \citep[see][]{rho18sn}; and (3) \texttt{pypeit} \citep{pypeit:joss_pub, pypeit:zenodo}. A discussion comparing the resulting spectra is left in the Appendix. The data reduction method using \texttt{pypeit} is described as well.

The reduction procedures of the semi-automated pipelines \texttt{Spextool} and \texttt{xdgnirs} are highly similar. Both ingested the science data, flats, and arcs obtained during the observation of \rbc\ and performed flat-fielding, wavelength calibration, aperture identification, tracing, and spectral extraction. For \texttt{Spextool}, flux calibration, order stitching, and telluric correction were carried out as separate, manually initiated steps following the initial automatic reduction. Telluric corrections were applied using observations of nearby standard stars (typically A0 stars) obtained on the same night as the science observations to minimize differences in airmass and atmospheric conditions.

The final reduced spectra are shown in \autoref{Figure:NIRSpectra}. The \texttt{xdgnirs} reduction of the GNIRS data is displayed in the figure; like the NIRES spectrum, this spectrum was smoothed using a third-order Savitzky–Golay filter with a 19-pixel window to improve feature visibility. Regions shaded in gray indicate portions of the spectra that are unreliable due to low atmospheric transmission.

\section{Results}\label{Section: Results}
\subsection{Explosion Date}\label{Subsection: Explosion Date}

Based on the photometry of \rbc\ available on TNS, the last non-detections reported by ZTF were on 2024 July 30 at 09:23:13 UTC in the $g$ band and at 10:18:01 UTC in the $r$ band ($m_{\text{Limit,AB}} = 20.50$ and $20.37$ mag, respectively). The first detection by ZTF was on 2024 August 3 at 08:20:55 UTC ($m_{\text{Limit,AB}} = 20.64$ mag in the $g$ band).
Additionally, ATLAS reported its last non-detection on 2024 August 3 at 13:23:19 UTC, but at a brighter limiting magnitude ($m_{\text{Limit,AB}} = 19.6$ mag in the orange ATLAS band). ATLAS reported its first detection at 12:08:20 UTC on 2024 August 4 ($m_{\text{Limit,AB}} = 19.4$ in the same band). 
Furthermore, the last non-detection by the Large Array Survey Telescope (LAST) was on 2024 August 2 at 22:40:06 UTC ($m_{\text{Limit, AB}} = 20.84$ mag, unfiltered). This constrains the window of the explosion date to between 22:40:06 UTC on August 2 and 08:20:55 UTC on August 3. In this work, we use an explosion date ($t_0$) of 03:30:31 UTC on 2024 August 3 (MJD $60525.15 \pm 0.20$), the midpoint between these two times. The epochs of our spectra are calculated with respect to this explosion date.

\subsection{Extinction}\label{Subsection: Extinction}

The extinction contributed by the Milky Way was estimated using the Galactic dust model developed by \cite{2011ApJ...737..103S}. We derived a reddening of $\text{E(B}-\text{V)}_{\text{Galactic}}= 0.0619 \pm 0.0011$ mag \citep[$R_V=3.1$;][]{fitzpatrick99}. To estimate the extinction from the host galaxy NGC 39, we examined the spectra for the optical sodium doublet (Na I D). No evidence of the doublet above the noise threshold was found in any of the optical spectra. Using the empirical relationship found by \cite{2012MNRAS.426.1465P} between the absorption strength of this line and extinction, we estimate $\text{E(B}-\text{V)}_{\text{Host}} < 0.02$ mag. The absence of evidence for significant host attenuation compels us to neglect the host's contribution. The extinction values are listed in \autoref{Table:KeyParameters}. 

\subsection{Light Curves}\label{Subsection: Light Curves}

The light curves of \rbc\ in the $BgcVroi$ bands are shown in \autoref{Figure:Photometry}. The $gro$ bands are relatively well sampled across the first 40 days. These light curves rise rapidly for the first $12-18$ days ($-0.18$ to $-0.27$ mag d$^{-1}$). Each band reaches a single, clear peak before declining quickly. The $g$ band peaks at $\sim$$12$ days, and the $ro$ bands at $\sim$$18$ days. This trend is extended to the other bands; bluer light curves peak earlier and decline more sharply afterward. 

Post-peak, the light curves undergo their first dimming phase between $20$ and $40$ days post-explosion, reaching $0.041$ to $0.055$ mag d$^{-1}$ in the \emph{gro} bands. Subsequently, between $40$ and $60$ days, the dimming rate transitions and slows. This transition noticeably smoother in redder bands; the $g$ band exhibits a sharp change between $40-45$ days, whereas the $r$ and $o$ bands do not. Past $60$ days, the dimming rate decelerates to between 0.014 and 0.017 mag d$^{-1}$. 

Generally, the light curves of the more sparsely sampled bands ($BcVi$) follow the trends described above. The $B$ and $V$ bands are best sampled around $\sim$40 days post-explosion and exhibit similar transitions in dimming rate. The $c$ band is the only non-$gro$ band to have pre-peak photometric data; \rbc\ peaks in the $c$ band later than the $g$ band. Lastly, the $i$ band shows a dimming rate transition that is highly similar to the $ro$ bands. For further analysis, the $r$ band light curve is compared to those of other SNe in Section \ref{Subsection: Light Curve Comparison}.

\subsection{Bolometric Luminosity and Light Curve Fitting} \label{Subsection: Bolometric Luminosity and Light Curve Fitting}

\begin{figure}
\centering
\includegraphics[scale=0.19]{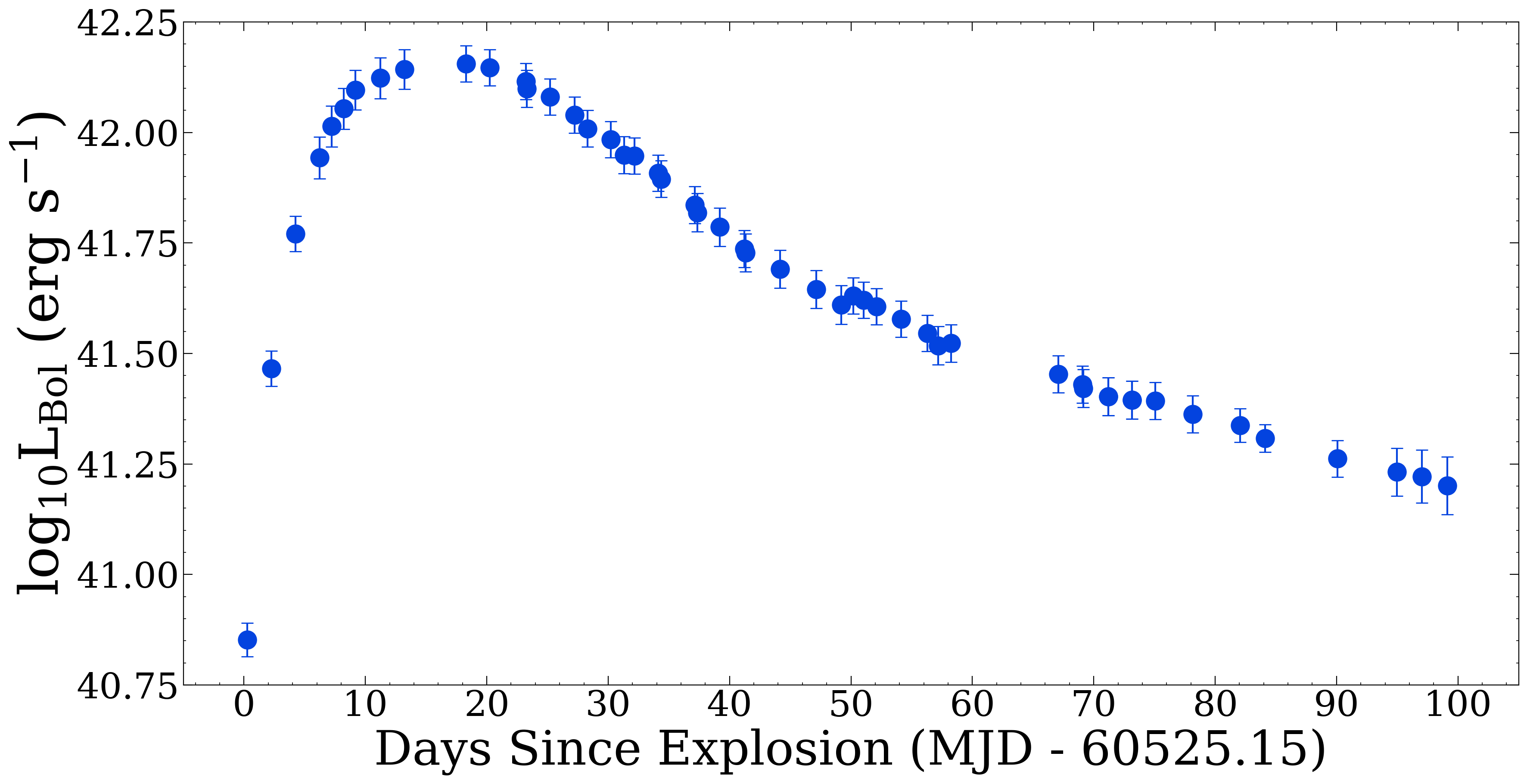}
\caption{\new{The quasi-bolometric light curve of \rbc\ as derived by \texttt{SuperBol}.}}
\label{Figure:BolometricLuminosity}
\end{figure}

Estimating the bolometric luminosity of a SN is essential to constraining the parameters of its progenitor. For this task, we used \texttt{SuperBol} \citep{2018RNAAS...2..230N}. This program takes in photometric data, \new{interpolates the observations and extrapolates across missing epochs, and numerically integrates observed flux to derive a quasi-bolometric luminosity. This derived L$_{\text{Bol}}$ represents the total flux measured across the observed wavelength bands, of which first 100 days post-explosion is displayed in \autoref{Figure:BolometricLuminosity}}. 

\begin{figure}
\centering
\hspace{-0.4cm}
\includegraphics[scale=0.25]{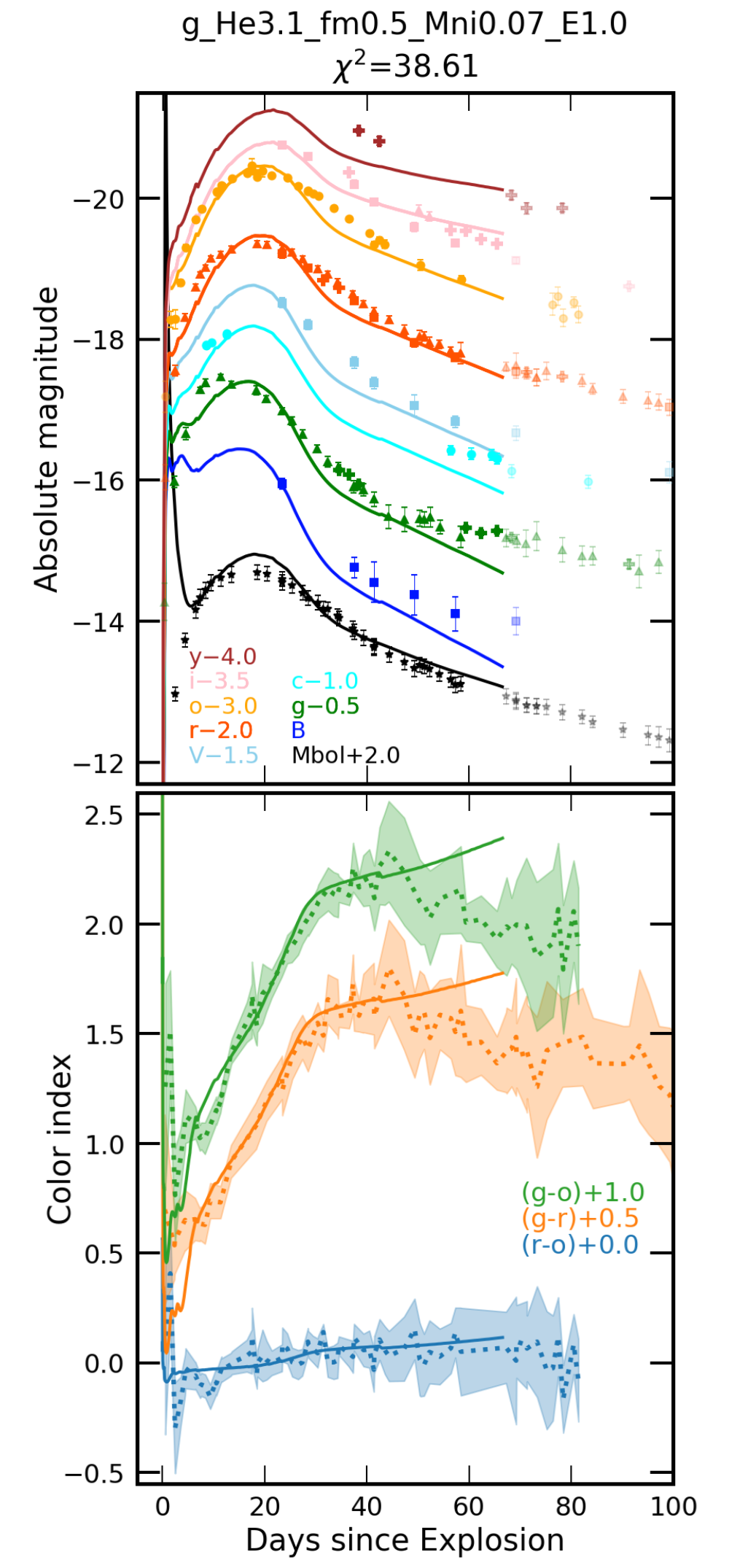} 
\caption{
\textit{Top}: Light curves of a $3.1$ $M_{\odot}$ He star progenitor Type Ib SN model superimposed on the photometry of \rbc. ZTF data are in triangles, ATLAS in circles, Nickel in squares\new{, and Pan-STARRS in "+" markers}; stars mark the bolometric magnitudes. Semi-transparent data points were not used when fitting the model.
The M$_{\text{Bol}}$ light curve is the same as \autoref{Figure:BolometricLuminosity}. The early peak in M$_{\text{Bol}}$ of the model is due to neglecting shock cooling \citep{jin23}. 
\textit{Bottom}: Color evolution in the $gro$ bands of \rbc\ (dotted) and the best-fit model (solid). Shaded regions about the data represent the uncertainty in color magnitude. \new{The units of both sub-figures are in magnitudes}}
\label{Figure:PhotometryFitting}
\end{figure}

These observed and calculated light curves were compared to Type Ib SN light curve models from the Type Ibc model grid presented in \citet{jin23}. These light curves were calculated via the 1D multi-group, radiative-hydrodynamical code STELLA \citep{blinnikov00, blinnikov06}, 
\new{which are based on the He star progenitor models computed under different assumptions, including models with/without binary companions and varying wind mass-loss rates \citep[see Section 3.1 of][for details]{jin23}. The varied parameters of the binary include the initial primary mass (ranging from 11 to 15 $M_\odot$), initial mass ratio (0.9), and initial orbital period (ranging from 50 to 200 days). The single star parameters include initial helium star masses ranging from $4$ to $9$ $M_\odot$.} Multi-color light curves are obtained by convolving the filter response functions with the SEDs computed at each time step.

The Type Ib SN models coarsely cover a wide range of values for each progenitor and explosion parameter. Six different ejecta masses ($1.7$ $M_\odot$ to $4.1$ $M_\odot$), four $^{56}$Ni masses ($0.07$ $M_\odot$ to $0.25$ $M_\odot$), two explosion energies (1 B and 2 B, where $1\,\text{B}=10^{51}$ erg), and six different $^{56}$Ni distributions (a step function and a Gaussian function, each with three different mixing parameters $f_\mathrm{m}$; see \citealt{yoon19} and \citealt{jin23} for its definition) were considered. As the resolution of the model grid is relatively coarse, any best-fit model should be interpreted as indicative rather than definitive. In other words, the best-fit model simply refers to the closest match within the model grid and is intended to provide approximate estimates rather than precise determinations of supernova parameters.

We fitted each Type Ib SN model's bolometric and multi-band light curves to those of \rbc\ and evaluated their goodness-of-fit using reduced $\chi^2$ values.

We note that STELLA assumes local thermodynamic equilibrium (LTE) for level populations, which is invalid past $\sim$$40$ days post-explosion. Therefore, the model light curves beyond this epoch should be considered with caution. \new{The fitting was performed from $1-65$ days post-explosion ($\sim$160 data points),
which corresponds to the time span covered by the STELLA simulations where the assumed circumstellar matter has little effect.} In the top panel of \autoref{Figure:PhotometryFitting}, we present the best-fit model light curves based on the reduced $\chi^2$ fitting. We note that restricting the fit to the first 40 days yields the same best-fit model as an unrestricted fit. The semi-transparent data points in \autoref{Figure:PhotometryFitting} show those masked during fitting.

The best-fit model for \rbc\ ("g\_He3.1\_fm0.5\_\\Mni0.07\_E1.0") represents a progenitor He star of $3.1$ $M_\odot$, with a $^{56}$Ni mass of $0.07$ $M_\odot$ and an explosion energy of 1 B. The $^{56}$Ni mixing is described by a Gaussian function with $f_\mathrm{m}=0.5$. The ejecta mass of this model is $1.7$ $M_\odot$. The progenitor corresponds to the "Sm11p200" model from \citet{yoon17}, which has an initial progenitor mass of $11$ $M_\odot$, mass ratio of 0.9, and orbital period of 200 days. \new{This model was selected from a grid of models with a final mass range of 3.1 to 5.6 $M_{\odot}$, corresponding to an initial mass range of 11 to 18 $M_{\odot}$ for primary stars in close binary systems with initial orbital periods of 20 to 500 days.}

The bottom panel of \autoref{Figure:PhotometryFitting} shows the color evolution in the $gro$ filters, which have good coverage at early times and around peak brightness. Since the photometric data were not collected simultaneously across different bands, we interpolated the light curves to estimate the observed colors. The color evolution of the model is broadly consistent with the observations until $\sim$40 days. At early times ($\lesssim 10$ days), \rbc\ exhibits blueward evolution followed by redward evolution, most noticeably in the $\rm g-o$ color. A qualitatively similar color evolution was observed in SN\,1999ex, another Type Ib SN (see Figure 11 in \citealt{yoon19}).

\rbc\ does not show a strong optical peak at early times ($\lesssim 5$ days post-explosion), indicating the absence of significant circumstellar material (CSM). Our best-fit model light curves were constructed using a progenitor model that assumes a moderate amount of CSM ($\sim$$0.018$ $M_\odot$). However, the CSM–ejecta interaction depends on factors such as the spatial extent and structural properties of the CSM \citep{piro15, jin21, khatami24, chiba25}, which are not included in our models. A full treatment of these effects is beyond the scope of this paper. The parameters derived from this fitting are listed in \autoref{Table:KeyParameters}.

\subsection{Optical Spectra}\label{Subsection: Optical Spectra Analysis}

The optical spectra of \rbc\ and the atomic lines we have identified are shown in \autoref{Figure:OpticalSpectra}. A variety of emission and absorption features are visible. Some of the strongest features exhibit identifiable P-Cygni profiles: the three Ca II lines at 8498, 8542, and 8662 \AA\ (which we refer to collectively as the Ca II triplet at 8567 \AA\ due to the high ejecta velocity causing overlap), O I at 7775 \AA, and He I at 5876 \AA. Other contributing species include Fe II, Si II, and C I. The optical sodium doublet (Na I D) is also marked, but it was not distinguishable in our spectra.

The Ca II absorption feature at 8567 \AA\ is contaminated by two narrower lines at approximately 8120 \AA\ and 8250 \AA. Both are visible at all epochs, and are weak compared to the strength of the Ca II absorption. These lines are likely telluric or galactic in origin and are not relevant to our analysis. Similarly, the O I absorption profile at 7775 \AA\ is affected by two narrow, irrelevant features that first appear in the 12 days post-explosion ($+12$d) spectrum and persist until at least $+41$d. The He I absorption feature at 5876 \AA\ overlaps with an unidentified broad emission feature near 5600 \AA. This overlap causes the He I feature to appear as two separate absorption minima beyond 12 days. While Ba II is a possible candidate for this emission, it is unlikely that barium would be present in sufficient quantities or have the broad velocity profile required.

Several line profiles exhibit significant temporal evolution. The C I lines at 9183 and 9406 \AA\ initially appear as absorption at +12d and transition to emission by $+24$d. All identified Ca II features also grow stronger over time. In particular, the Ca II line at 7291 \AA\ appears to show a P-Cygni profile at $+61$d. Although it partially overlaps with a He I line, the pronounced increase in strength at later epochs indicates that this feature is dominated by Ca II. Likewise, the identified He I and O I features also grow in prominence relative to the continuum.

Another noticeable trend is the strengthening of the Fe~II lines below 6000 \AA. The shortest-wavelength Fe~II line (4233 \AA) is clearly detected from $+6$ to $+36$d, after which it falls below the noise threshold. The other three Fe II lines (5169, 5363, and 5535 \AA) grow stronger as \rbc\ ages, contributing to the gradually rising, continuum-like shape of the spectrum in the region.

\subsection{Near-Infrared Spectra}\label{Subsection: NIR Spectra Analysis}

\autoref{Figure:NIRSpectra} displays the NIR spectra at 40, 47, and 62 days post-peak, along with the spectral lines we have identified. Unlike the optical spectra, the NIR observations are limited to epochs after the peak and first decline phase ($\geq40$ days).

In the NIR, neutral atoms dominate the spectra, whereas ionized atoms dominate in the optical. Species producing strong lines in the NIR include C I, S I, Mg I, Si I, Fe II, Na I, and Ca I. Neutral helium and oxygen also contribute. The He I P-Cygni profile at 1.083 $\mu$m is the strongest feature, analogous to the Ca II triplet in the optical spectra. However, the prominence of this profile makes it difficult to discern weaker lines in its vicinity. The small dip at $\sim$$1.0$ $\mu$m, just short of the He I absorption, resembles the Sr II line identified by \citet{dong23}.

Several emission lines increase in prominence over time.  An emission feature of C I appears at 0.918 $\mu$m, first visible at +47d, and becomes stronger by +62d. This is the same C I feature that was observed in the optical spectra. The O I emission line at 1.129 $\mu$m evolves in a similar manner. Other lines showing relative growth in emission strength against the continuum include Mg I at 1.504 and 1.711 $\mu$m, Si I at 1.589 $\mu$m, and Na I at 2.206 $\mu$m. The particularly broad Mg I line at 1.504 $\mu$m, which maintains its width over time, suggests that magnesium has retained most of its initial velocity from the explosion along our line of sight.

A small portion of the optical spectrum was measured by SpeX 47 days post-explosion. The strong Ca II triplet and O I features are clearly evident in this spectrum. Additionally, GNIRS captured the emission peak of the Ca II P-Cygni profile at +62d.

A He I line at 2.058 $\mu$m is also visible. This line forms a clear P-Cygni profile, most prominent at +40d. While another He I line may exist at 2.112 $\mu$m, it is weak relative to the 2.058 $\mu$m line and no attributable feature is discernible. Furthermore, the absorption profile of the 2.058 $\mu$m line is heavily contaminated by a strong CO$_2$ telluric absorption band near $\sim$$2$ $\mu$m. This contamination makes it difficult to determine the detailed structure of the 2.058 $\mu$m He I feature beyond the broad P-Cygni profile present in all spectra.

Finally, we detect the first CO overtone in the +62d spectrum. Typically, this overtone exhibits band heads at 2.294, 2.323, 2.353, 2.383, 2.414, and 2.446 $\mu$m. The rise in emission beyond $\sim$$2.27$ $\mu$m marks the onset of the band head. The high opacity and velocity broadening of the CO make the individual band heads indistinguishable. Additionally, the slightly rising continuum at $\sim$$2$ $\mu$m indicates the presence of warm dust. The results of modeling these features are discussed in Section \ref{Subsection: CO and Dust Modeling}.

\section{Discussion}\label{Section: Discussion}
\subsection{Light Curve Comparison}\label{Subsection: Light Curve Comparison}

\begin{figure*}
\centering
\includegraphics[scale=0.5]{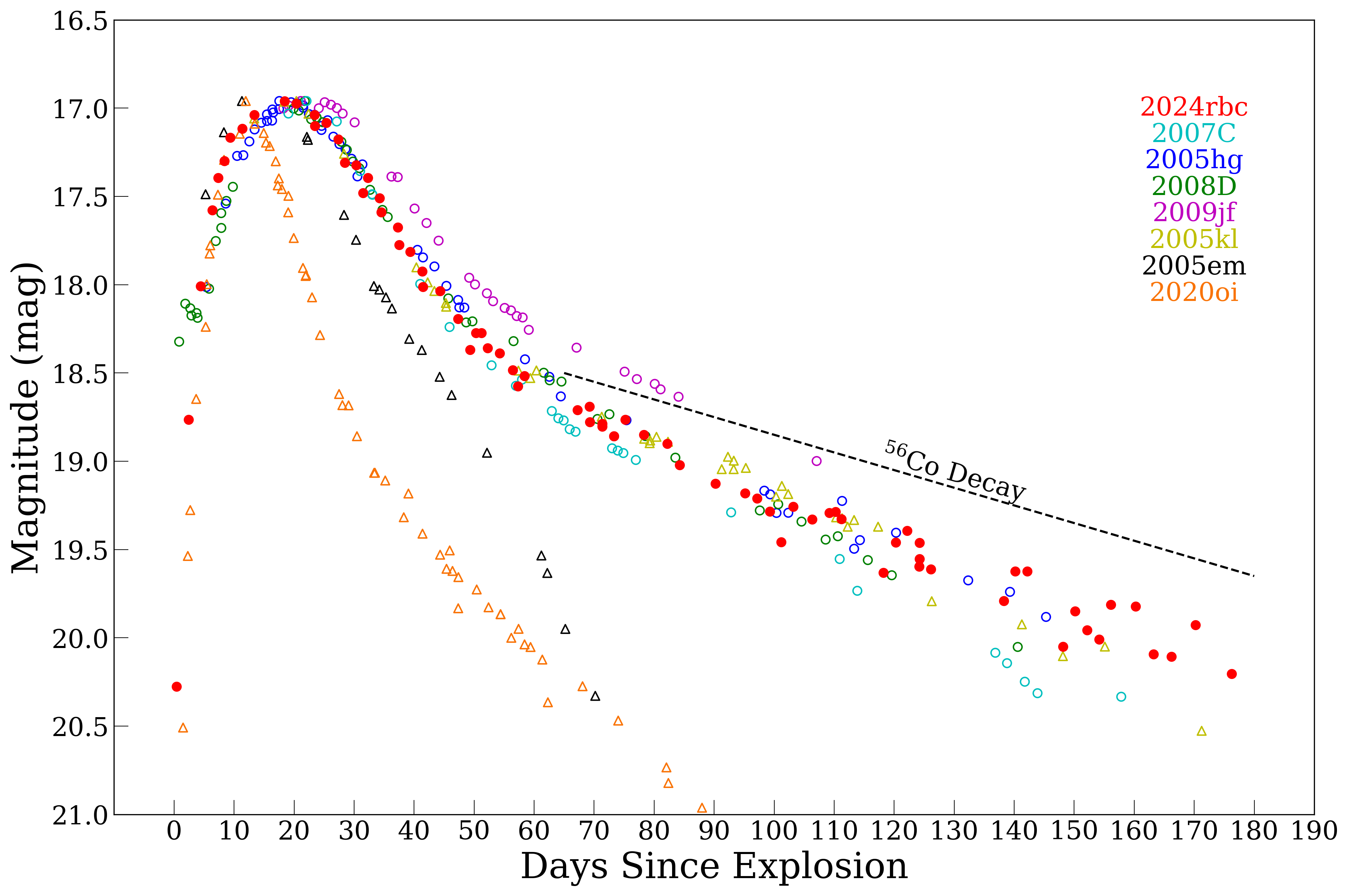}
\caption{A comparison of the \emph{r} band light curve of \rbc\ (solid red circles) against other Type Ib and Ic SNe. SNe 2007C, 2005hg, 2008D, and 2009jf are Type Ib (open circles). SNe 2005kl, 2005em, and 2020oi are Type Ic (open triangles). The magnitudes of each light curve have been adjusted such that the peak matches that of \rbc.}
\label{Figure:PhotometryComparison}
\end{figure*}

The \emph{r} band light curve of \rbc\ is compared Type Ic SNe 2020oi \citep{rho21}, 2005em \citep{sako18}, and 2005kl \citep{bianco14} and Type Ib SNe 2007C, 2005hg, 2008D, and 2009jf \citep{bianco14} in \autoref{Figure:PhotometryComparison}. The decay rate expected for fully trapping the $^{56}$Co decay emission (0.0098 mag d$^{-1}$) is illustrated for $t>65$ days as well. The light curve data for these SNe were sourced from the Open Supernova Catalog\footnote{\url{https://github.com/astrocatalogs/supernovae}} \citep[OSC;][]{OSC}.

The very early rise of \rbc\ is similar to that of SN\,2020oi. SNe 2020oi and 2005em peak far earlier and more sharply than the rest of the SNe in this sample, in addition to declining rapidly afterwards. This indicates that \rbc\ significantly diverges from some, but not all, Type Ic SNe. 

Close to the peak, the rising portion of \rbc's light curve resembles SNe 2007C and 2008D, but \rbc\ rises slightly higher before peaking. The peak of \rbc\ lines up remarkably well with SNe 2007C, 2005hg, 2008D, and 2005kl; all of these SNe peak at $\sim$$18$ days. Of the Type Ib SNe, only SN\,2009jf seems to be delayed, peaking $>20$ days. Post-peak, the first decline is highly similar between \rbc\ and the Type Ib SNe. From 18 to 50 days, these SNe decline at a rate of $\sim$$0.44$ mag d$^{-1}$. 

Afterwards, from 50 to 65 days, the light curves transition to a second dimming rate. \rbc\ does not trap the entirely of the $^{56}$Co decay emission, much like all the other SNe. From 65 to 160 days, \rbc\ declines at a rate of 0.0137 mag d$^{-1}$. This rate is matched or exceeded by all of the other sampled Type Ib SNe, including SN\,2009jf. 

This epoch is also where the light curve of \rbc\ becomes less consistent with that of SN\,2007C. The latter declines more quickly past 65 days than the other SNe. After $\sim$$125$ days, SN\,2005kl dims faster than \rbc\ as well. Considering that uncertainties in photometric data grow as SNe dim, we conclude that the decay rates and light curve morphologies of SNe 2024rbc, 2005hg, and 2008D are largely consistent. This agreement supports \rbc\ being Type Ib, but does not rule out a Type Ic classification.

\subsection{Comparison of Optical Spectral Evolution}\label{Subsection: Optical Spectra Evolution and Comparison}

\begin{figure*}
\centering
\includegraphics[scale=0.36]{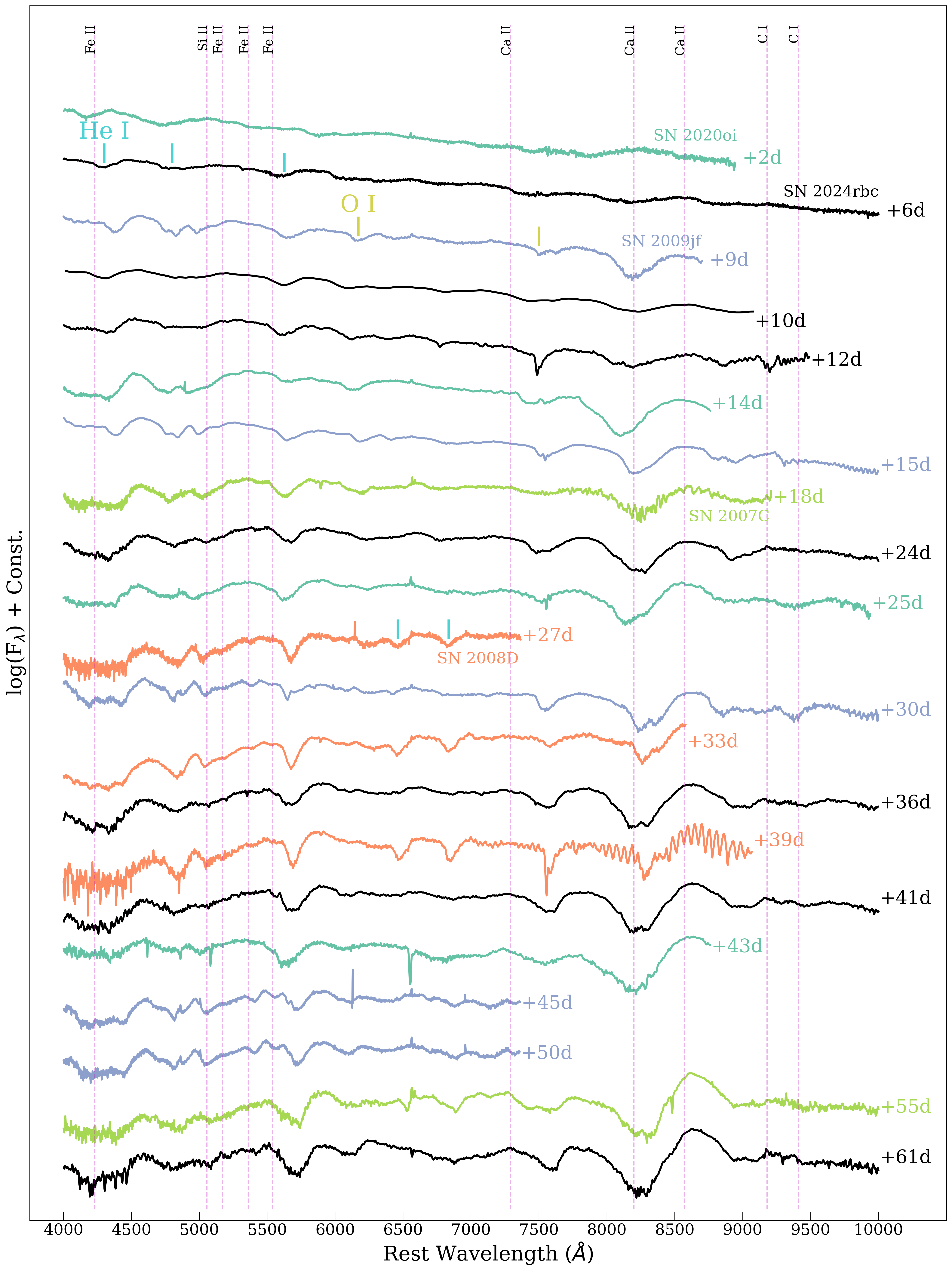}
\caption{An array of optical spectra from \rbc\ and several Type Ib and Ic SNe. \rbc\ is in black. SNe 2009jf, 2007C, and 2008D are Type Ib (marked in gray-blue, lime, and orange, respectively). SN\,2020oi is Type Ic (marked in teal). Key metal lines are marked by the dashed magenta lines. Neutral helium and oxygen features are marked individually in cyan and yellow, respectively, at the absorption minimum.}
\label{Figure:OpticalComparison}
\end{figure*}

\autoref{Figure:OpticalComparison} shows optical spectra from five supernovae. The spectra for the comparison SNe, except SN\,2020oi, were sourced from OSC. By comparing the spectra of \rbc\ with those of several Type Ib and Ic SNe, we illustrate \rbc's consistency with the spectral features and evolution characteristic of Type Ib SNe.

Less than a week after the explosion, the spectrum of \rbc\ exhibits a relatively smooth, blackbody-like profile, similar to that seen in SN\,2020oi \citep{rho21}. As the peak luminosity is reached and passed (around $10-20$ days post-explosion), the spectra flatten, and spectral features develop rapidly in all sampled spectra.

The spectroscopic distinction between Type Ib and Type Ic SNe is made by He I features. The shorter He I lines ($<6000$ \AA) are visible to varying degrees in all sampled supernovae, including \rbc. He I lines at longer wavelengths ($\sim$$6500$ and $6800$ \AA), which are most clearly seen in the +27d spectrum of SN\,2008D \citep{richardson01, malesani09, moskvitin10, yaron2012wiserep, modjaz14}, are also visible in \rbc\ (most evidently in the +12d spectrum). However, these weaker lines near are absent in the Type Ic SN\,2020oi.

Additionally, spectral features from metals in \rbc\ show similarities to those of other SNe. The marked Ca II features are similar across all sampled spectra and become apparent at least as early as 9 days post-explosion. The O I feature at 7775 \AA\ is also visible in all spectra. Furthermore, the growing prominence of Fe II emission lines below 6000 \AA, which form a continuum-like structure in \rbc, is also observed in the $>25$ day spectra of SN\,2008D and SN\,2007C \citep{silverman12, shivvers19}. SN\,2009jf \citep{valenti11, silverman12, yaron2012wiserep, modjaz14, shivvers19} exhibits these Fe II lines as well, although they appear more distinct.
The C I lines, while clearly visible in \rbc, are not apparent in any of the other spectra, with the possible exception of SN\,2009jf.

Based on these spectral similarities, we draw the same conclusion as from the light-curve comparisons: the optical spectra of \rbc\ and their evolution are most consistent with \rbc\ being a Type Ib SN.

\subsection{Comparison of NIR Spectral Evolution}\label{Subsection: NIR Spectra Evolution and Comparison}

\begin{figure*}
\centering
\includegraphics[scale=0.36]{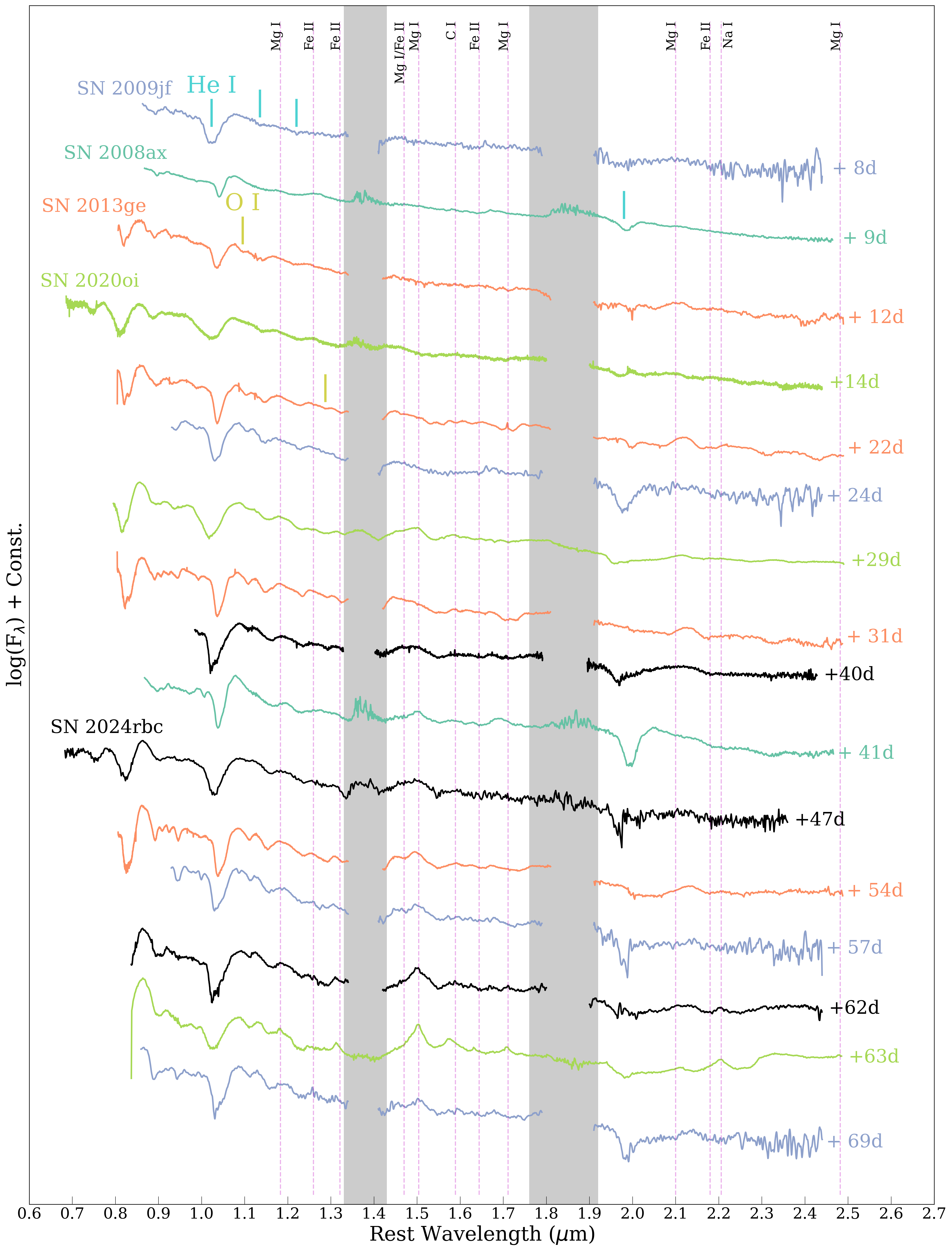}
\caption{An array of NIR spectra from \rbc\ and several Type IIb, Ib, and Ic SNe. \rbc\ is marked in black. SN\,2009jf is Type Ib and SN\,2008ax is Type IIb (marked in gray-blue and turquoise, respectively). SNe 2020oi and 2013ge are Type Ic (marked in lime and orange, respectively). Key metal lines are marked in dashed magenta. Neutral helium and oxygen are marked individually at the absorption minimum in cyan and yellow, respectively. Regions of high uncertainty due to low atmospheric transmission are marked in gray.} 
\label{Figure:NIRComparison}
\end{figure*}

\autoref{Figure:NIRComparison} compares the NIR spectra of \rbc\ to four other SNe. Again, the spectra for the comparison SNe, except SN\,2020oi, were sourced from OSC. The strongest features in the spectra are the He I lines at 1.083 and 2.058 $\mu$m. A P-Cygni profile is observed at the former in all sampled spectra. The asymmetric absorption profile seen in \rbc\ is also visible in SN\,2013ge \citep{drout16, yaron2012wiserep} and SN\,2009jf \citep{valenti11, yaron2012wiserep}, whereas SN\,2020oi does not exhibit this asymmetry. Additionally, the P-Cygni profile at 2.058 $\mu$m is clearly visible in SNe 2009jf and 2008ax \citep{taubenberger11, yaron2012wiserep}, but is much weaker in SNe 2020oi and 2013ge. Interestingly, the +57d profile of SN\,2013ge is very similar to that of \rbc\ at +47d, and the aforementioned line contamination also appears to be present.

Weaker He I features at 1.197 and 1.278 $\mu$m overlap with metal lines. The neighboring Mg I and Fe II emission lines make it difficult to distinguish the He I lines past $\sim$$20$ days post-explosion. By $\gtrsim 50$ days, the metal lines dominate all sampled spectra at these wavelengths.

The O I emission line adjacent to the He I feature at 1.083 $\mu$m first appears in the +12d spectrum of SN\,2013ge. This feature is absent only in the spectra of SN\,2008ax. Similarly, the Na I emission line at 2.206 $\mu$m is observed only in \rbc\ and SN\,2020oi \citep{rho21}. The strong Mg I line at 1.504 $\mu$m is evident in all sampled spectra past +40d.

Lastly, the first CO overtone and a dust continuum are observed in the +62d spectrum of \rbc. A more pronounced version of this feature is visible in the +63d spectrum of SN\,2020oi \citep{rho21}. Comparisons with SNe 2009jf and 2013ge highlight the atypical nature of this feature; in the absence of warm dust and CO, the spectrum would be expected to be nearly flat or slightly declining. Further analysis of this feature is presented in Section \ref{Subsection: CO and Dust Modeling}.

\subsection{Line Analysis}\label{Subsection: Line Velocity Analysis}

\begin{figure}
\centering
\includegraphics[scale=0.4]{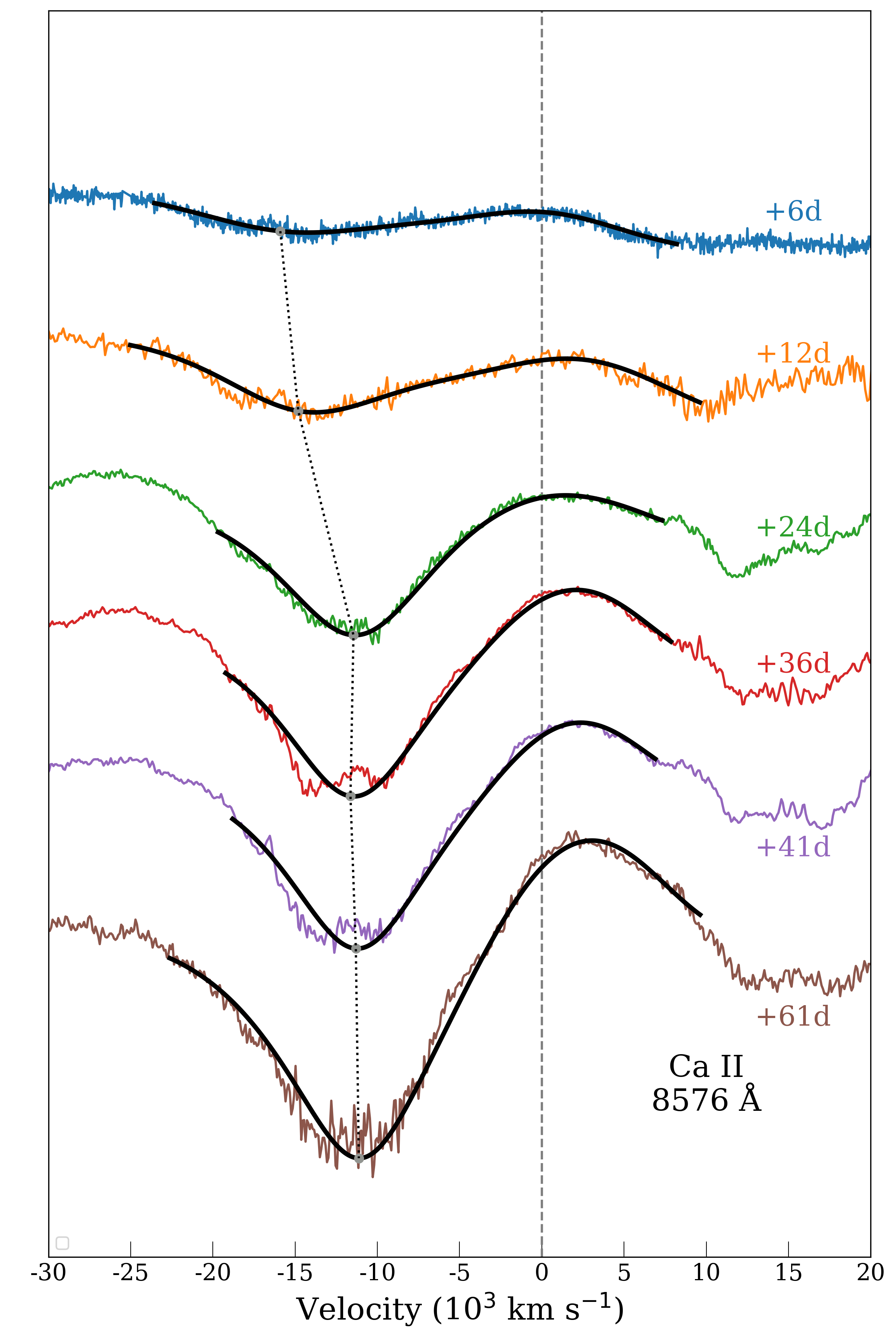}
\caption{The evolution of the Ca II P-Cygni profile at 8567 \AA, fitted by a two-Gaussian approximation of the profile. The black lines indicate the fitted profiles, and the gray dotted lines connect the absorption minima. A gray dashed line has been drawn at rest as a visual aid.}
\label{Figure:CaIIPCygni}
\end{figure}

To better understand the evolution of \rbc, we fitted a two-Gaussian approximation of a P-Cygni profile to the NIR Ca II triplet at $\sim$8567 \AA. The best-fit profiles and corresponding absorption minima for each optical spectrum are shown in \autoref{Figure:CaIIPCygni}. The +10d spectrum was omitted from this analysis due to low SNR. All fitted parameters are listed in \autoref{Table:CaIIPCygni}.

The evolution follows an intuitive progression. The absorption minimum has an initial velocity $>1.3\times10^{4}$ \kms\ (blueshifted), which rapidly decelerates during the first $20-25$ days. After this initial deceleration, the velocity plateaus to $8.5-9.0\times10^{3}$ \kms\ by +24d. The center of the P-Cygni profile shifts redward over time, starting at $-4\times10^{3}$ \kms\ at +6d and ending at $\sim$300 \kms\ at +61d.

Relative to the absorption, the emission maximum exhibits a muted temporal evolution. Starting near rest, the maximum becomes increasingly redshifted and settles at $\sim$$3\times10^{3}$ \kms\ by +36d. This suggests that the blueshifted Ca II in \rbc\ has decelerated significantly during the first two months, while the redshifted portion has accelerated away from our line of sight.

Additionally, the FWHM of the absorption and emission components (see \autoref{Table:CaIIPCygni}) illustrate that the blueshifted absorption of Ca II becomes broader and more prominent as the SN cools. This is consistent with expectations: the blueshifted Ca II, along with other ionized metals in the ejecta and CSM, becomes a dominant spectral feature in the optical spectra as the continuum weakens. The emission profile, in contrast, is widest near peak luminosity, suggesting that the expanding ejecta encounters resistance as it interacts with surrounding material.

\begin{figure*}
\centering
\includegraphics[scale=0.275]{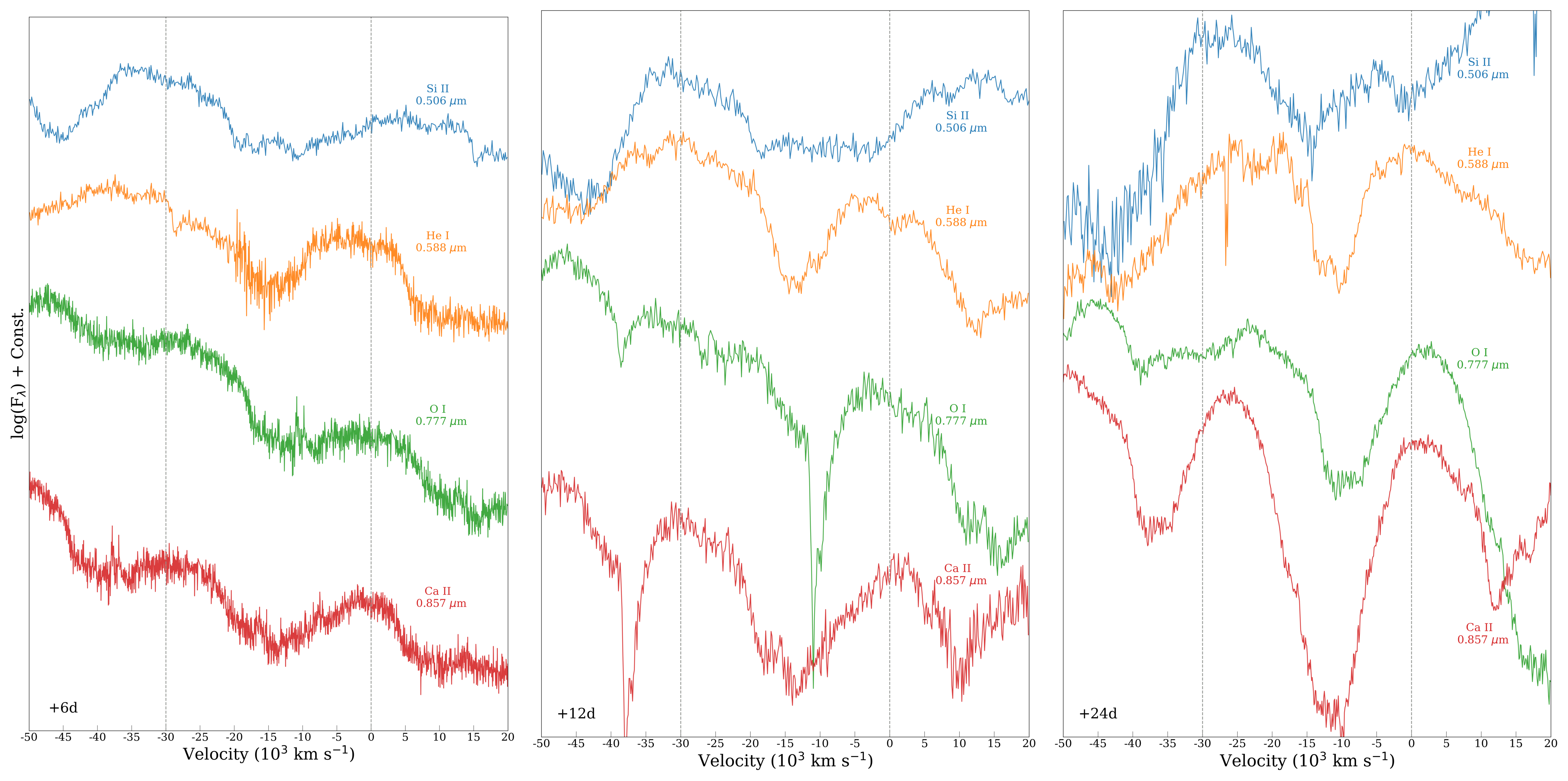}
\caption{A comparison of the velocity profiles of key lines (Si II, He I, O I, and Ca II) in the optical regime at +6d, +12d, and +24d. The two vertical, dashed lines in gray indicate the velocity range that contains the features of interest.}
\label{Figure:MetalVelocityComparison}
\end{figure*}

We also compare the velocities of different metal absorption features in \autoref{Figure:MetalVelocityComparison}. At +6d, the absorption features are fairly well aligned with minima near $-1.5\times10^{4}$ \kms. The O I minimum is slightly lower, likely due to a broader absorption profile. We also note that the Si II feature may be minorly contaminated by a weak He I line (5016 \AA) and at least one Fe II line (5169 \AA).

At +12d, the Si II absorption profile has broadened; the short-wavelength cut-on remains roughly constant, but the profile has extended toward longer wavelengths. The He I line has formed a sharper and clearer absorption profile and has maintained a minimum near $-1.5\times10^{4}$ \kms. Similarly, the Ca II profile has deepened without a significant change in velocity. In contrast, the O I line shows a reduced absorption minimum at approximately $-1.0\times10^{4}$ \kms, deviating from the other lines, and is contaminated by at least one sharp absorption feature. However, overall, the line velocities have not changed significantly between +6 and +12d.

Finally, at +24d, noticeable changes are observed in the each of the features. The He I and Ca II absorption profiles have decelerated, with minima at approximately $-1.2\times10^{4}$ \kms. The O I line has decelerated further to nearly $-1.0\times10^{4}$ \kms. The Si II line is increasingly contaminated, making its absorption profile difficult to ascertain at this epoch. By this time, the emission peaks of the lines are largely similar, with velocities exceeding $5\times10^{3}$ \kms\ relative to rest. This velocity evolution suggests that the ejecta of \rbc\ encountered some resistance between +12 and +24d, most likely from interacting with surrounding material.

\subsection{CO and Dust Detection and Modeling}\label{Subsection: CO and Dust Modeling}

\begin{figure}
\centering
\includegraphics[scale=0.4]{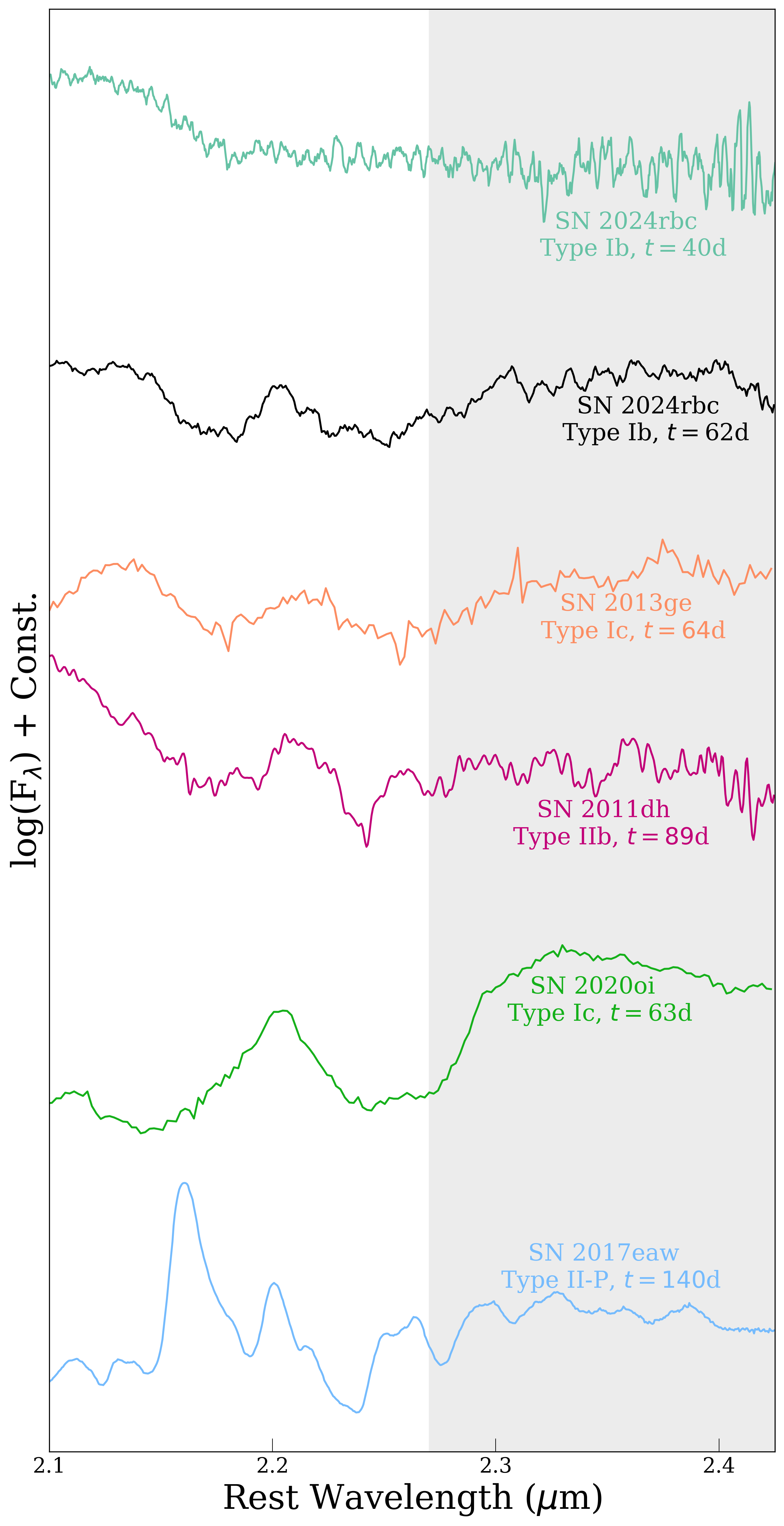}
\caption{A comparison of the CO-exhibiting NIR spectra of SNe 2013ge, 2011dh, 2020oi, and 2017eaw to the +40d and +62d spectrum of \rbc. The +40d spectrum serves as a comparison baseline for the other spectra. The gray region indicates the wavelength range of the first CO overtone.}
\label{Figure:COComparison}
\end{figure}

In \autoref{Figure:COComparison}, we compare the NIR spectra of SNe that have been shown to exhibit the first CO overtone to the +40d and +62d spectra of \rbc. The +40d spectrum of \rbc\ serves as a non-detection baseline, while SNe 2013ge \citep{drout16}, 2011dh \citep{ergon14, yaron2012wiserep}, 2020oi \citep{rho21}, and 2017eaw \citep{rho18sn} are shown for comparison to the +62d spectrum. The spectra for SNe 2013ge and 2011dh were sourced from OSC.

\rbc\ clearly exhibits elevated emission past $\sim$$2.27$ $\mu$m at +62d. The shape of the CO feature is distinct from those seen in Type II SNe 2011dh and 2017eaw, where several band heads are visible. Since CO is only detectable at later epochs in Type II SNe relative to Type Ib and Ic SNe, the lower CO velocities render the band heads distinguishable. 

Comparing \rbc\ with other Type Ib and Ic SNe, we see that the overtone in SN\,2020oi is much stronger. On the other hand, SN\,2013ge provides a good match as it is similar in profile shape and relative strength. The strong similarity suggests that this is a genuine detection of CO in \rbc.

\begin{figure*}
\centering
\includegraphics[width=13truecm]{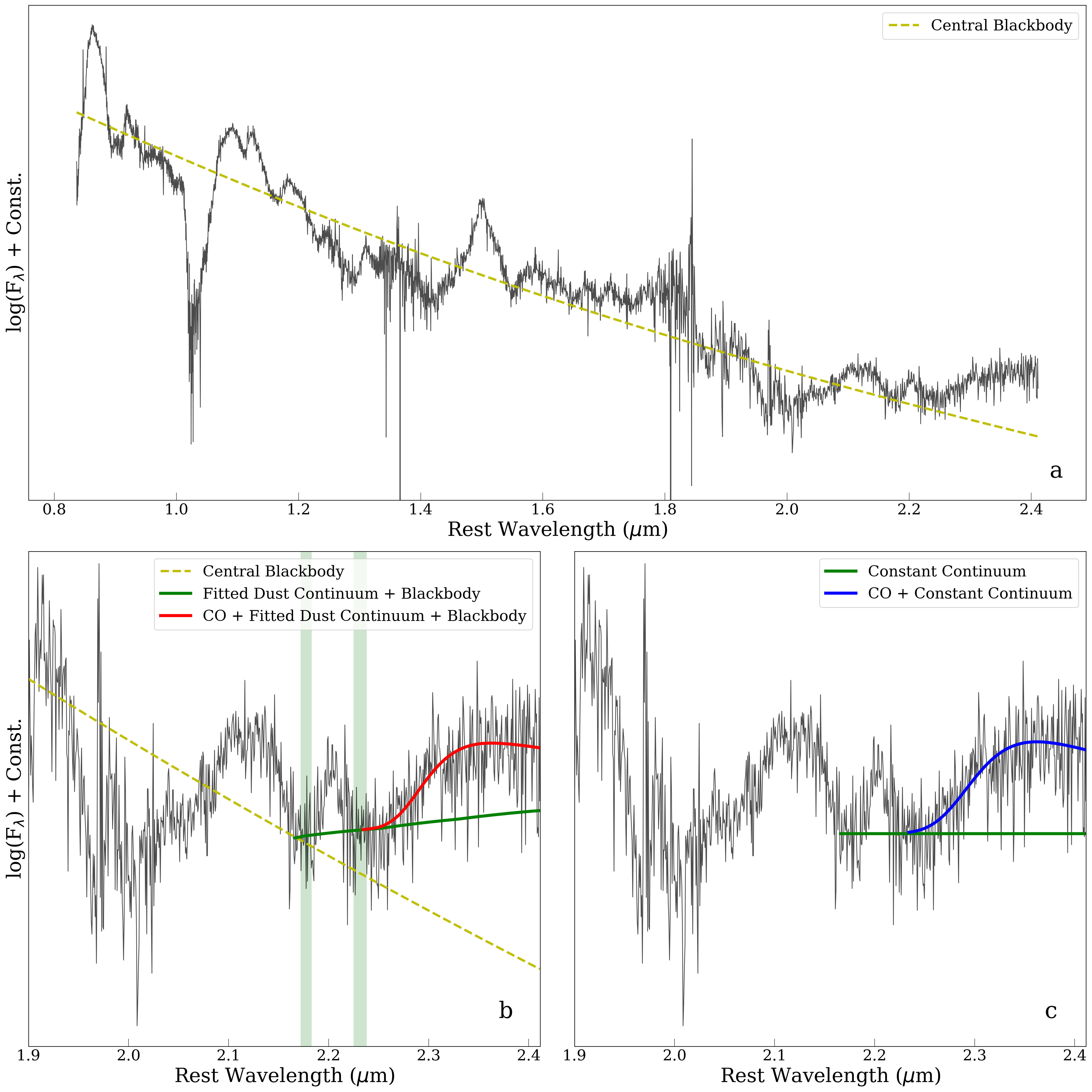}
\caption{\textit{Top} (\textit{a}): A view of the blackbody fit to the entire +62d NIR spectrum. \textit{Bottom Left} (\textit{b}): A zoomed-in view of the +62d NIR spectrum in the region of interest for the first CO overtone. The blackbody, dust continuum, and overtone fit are overlaid in dashed yellow, solid green, and solid red, respectively. The green regions indicate the portions of the spectrum used to fit the dust continuum. \textit{Bottom Right} (\textit{c}): \new{The same view as the left, but now with an alternate assumption of the continuum with the. The flat continuum is marked in green and the CO overtone fit in blue.}}
\label{Figure:COFit}
\end{figure*}

Modeling the CO emission requires an estimate of the continuum. Looking at the spectra as a whole (\autoref{Figure:COFit}a), it is clear that the continuum the spectrum rests upon is not the same between the region of the first CO overtone ($>2.27$ $\mu$m) and the rest of the spectrum. At shorter wavelengths, a simple blackbody spectrum fits the continuum. However, a greater continuum contribution exists just beyond the He I P Cygni profile at 2.06 $\mu$m. We argue that this additional continuum component is contributed by recently formed warm ($\lesssim 1100$ K) dust in the ejecta of \rbc. We expect this warm dust to contribute a rising K-band continuum, which the CO overtone would sit upon.

To account for these components, we fit the +62d NIR spectrum in three parts: the blackbody emission from the hot core of the SN, the continuum emission from the warm dust, and the emission from the CO. Here, we provide a description of the models for clarity and convenience. Section 3.4 in \citet{park25} also describes this model.

The blackbody emission from the photosphere was fit using a two-parameter Planck's Law model
\begin{equation}
\label{Equation:BBModel}
F_\lambda = C\frac{2hc^2}{\lambda^5}\frac{1}{e^{h c / \lambda k T} - 1},
\end{equation}
where $C$ is a fitting constant for flux scaling and $T$ is the effective temperature of the blackbody.

\cite{millard21} describes a method of modeling continuum emission from warm dust via a modified version of Planck's Law. This model is expressed as 
\begin{equation}
F_\lambda = \frac{M_{\text{dust}} \kappa_\lambda}{d^2} B_\lambda(T),
\end{equation}
where $F_\lambda$ is the flux at wavelength $\lambda$, $M_{\text{dust}}$ is the dust mass, $\kappa_\lambda$ is the wavelength-dependent dust absorption coefficient, $d$ is distance to the source, and $B_\lambda(T)$ is Planck's Law. The absorption coefficient can further be expressed as
\begin{equation}
\kappa_\lambda = \frac{3}{4 \rho a}Q_{\text{abs}},
\end{equation}
where $Q_{\text{abs}}$ is a wavelength-dependent absorption coefficient, $\rho$ is the density of carbon ($2.28 \times 10^{-3}$ g cm$^{-3}$), and $a$ is the grain size. Considering that $Q_{\text{abs}}$ is the only wavelength-dependent parameter, if we collect the constants and $M_{\text{dust}}$ needed to solve for $F_\lambda$ into a new, fittable parameter ($N_{\text{const}}$), we get
\begin{equation}
N_{\text{const}} = \frac{3}{4 \rho a d^2} M_{\text{dust}}.
\end{equation}
From $N_{\text{const}}$, we can estimate $M_{\text{dust}}$. Thus, the dust continuum emission can be modeled as
\begin{equation}
\label{Equation:DustModel}
F_\lambda = N_{\text{const}} Q_{\text{abs}} B_\lambda(T),
\end{equation}
leaving us with two parameters to fit for: $T$ and $N_{\text{const}}$. 

Lastly, the model for the CO flux is derived from \cite{cami10}, which assumes that the CO is isothermal and in LTE. We used line data from \cite{goorvitch94} to inform the model on the physical parameters of the CO molecules. We also assumed that the CO local to the supernova is composed of pure $^{12}$C$^{16}$O, which has been found to be adequate \citep[see][]{banerjee18, rho18sn, rho24, park25}. Equation (1) of \cite{cami10} calculates the line strength $S_{\nu_0}$ with respect to the transition frequency $\nu_0$ as:
\begin{equation}
S_{\nu_0} = \frac{h\nu_0}{4 \pi} g_1 B_{12} \frac{e^{-E_1/kT}}{P(T)} \left( 1 - e^{-h\nu_0/kT} \right).
\end{equation}
Combined with Equation 5 of \cite{goorvitch94}, we get
\begin{equation}
S_{\nu_0} = \left( 8.8523 \times 10^{-13} \right) \frac{gf}{P(T)} \frac{1 - e^{-h\nu_0/kT}}{e^{E_1/kT}} ,
\end{equation}
where $gf$ is the product of the statistical weight $g$ and emission oscillator strength $f$, $P(T)$ is the partition function, and $E_1$ is the lower level energy. The frequency-dependent CO flux is calculated by 
\begin{equation}
\label{Equation:COModel}
F_\nu = \frac{4 \pi \left( V_{\text{CO}} t \right)^2}{d^2} B_\nu \left( T_{\text{CO}} \right) \sum_{i} N_{\text{CO}} S_{\nu_0}^{i} \phi \left( \nu , V_{\text{CO}} \right),
\end{equation}
where $V_{\text{CO}}$ is the line velocity, $t$ is time since the explosion, $B_\nu$ is Planck's Law with respect to frequency, $N_{\text{CO}}$ is the column density, and $\phi \left( \nu , V_{\text{CO}} \right)$ is the Gaussian velocity broadening function that sets the line velocity as the FWHM. Using this model, we fit for the CO line velocity \new{width}, temperature, and dust mass $M_{\text{CO}}$, which is estimated from $N_{\text{CO}}$ as 
\begin{equation}
M_{\text{CO}} = 4 \pi \left( V_{\text{CO}} t \right)^2 m N_{\text{CO}},
\end{equation}
where $m$ is the molecular mass of CO ($\sim$$28$ AMU).

\begin{table*}
    \begin{center}
    \caption{NIR Ca II Triplet P-Cygni Fit Parameters}\label{Table:CaIIPCygni}
    \vspace{-0.4cm}
        \begin{tabular}{ccccccccccc}
         \toprule 
         Epoch  & \multicolumn{2}{c}{Minimum} & \multicolumn{2}{c}{Center} & \multicolumn{2}{c}{Maximum} & \multicolumn{2}{c}{Absorption FWHM} & \multicolumn{2}{c}{Emission FWHM} \\
         \cmidrule(lr){2-3} \cmidrule(l){4-5} \cmidrule(l){6-7} \cmidrule(l){8-9} \cmidrule(l){10-11}
         & $\lambda$ & $v$ & $\lambda$ & $v$ & $\lambda$ & $v$ & $\Delta\lambda$ & $\Delta v$ & $\Delta\lambda$ & $\Delta v$ \\
         & ($\mu$m) & ($10^{3}$ km s$^{-1}$) & ($\mu$m) & ($10^{3}$ km s$^{-1}$) & ($\mu$m) & ($10^{3}$ km s$^{-1}$) & ($\mu$m) & ($10^{3}$ km s$^{-1}$) & ($\mu$m) & ($10^{3}$ km s$^{-1}$) \\
         \midrule
          +6  & 0.812 & -13.4 & 0.838 & -4.1 & 0.859 & 3.0 & 0.034 & 11.8 & 0.026 &  9.1 \\
         +10  & 0.818 & -11.1 & 0.841 & -3.1 & 0.863 & 4.4 & 0.035 & 12.4 & 0.033 & 11.6 \\
         +12  & 0.815 & -12.3 & 0.841 & -3.1 & 0.865 & 5.3 & 0.032 & 11.4 & 0.029 & 10.3 \\
         +24  & 0.825 &  -8.9 & 0.843 & -2.3 & 0.861 & 3.9 & 0.034 & 11.9 & 0.032 & 11.4 \\
         +36  & 0.824 &  -9.1 & 0.846 & -1.3 & 0.864 & 4.8 & 0.033 & 11.5 & 0.025 &  8.9 \\
         +41  & 0.825 &  -8.7 & 0.848 & -0.6 & 0.864 & 4.9 & 0.035 & 12.2 & 0.024 &  8.4 \\
         +61  & 0.826 &  -8.5 & 0.851 &  0.3 & 0.866 & 5.5 & 0.044 & 15.4 & 0.025 &  8.9 \\
         \bottomrule
        \end{tabular}
    \end{center}
\end{table*}

\begin{table}
    \begin{center}
    \caption{CO \& Dust Parameters}
    \label{Table:COFitParametersNew}
    \resizebox{\columnwidth}{!}{
    \hspace{-1.35cm}
        \begin{tabular}{l|cc}
        \hline \hline 
        Parameter                        &  CO + Dust Cont. & CO + Flat Cont.  \\
        \hline
        CO Temp. (K)                     &      4040 (435) &  4170 (390)   \\ 
        
        CO Velocity \new{Width} (km s$^{-1}$)   &   5905 (1960) & 7970 (1620)    \\     
        CO Mass ($10^{-4}$ $M_{\odot}$)    &  5.2 (1.2)  &    6.1 (1.2)    \\
        \hline
        Dust Temp. (K)                   &    910 (10)   &  ...         \\     
        Dust Mass ($10^{-3}$ $M_{\odot}$) &   1.3 (0.1)   &    ...         \\ 
        \hline \hline 
        \end{tabular}}
        \end{center}
\renewcommand{\baselinestretch}{0.8}
\vspace{-7pt}
\footnotesize{The blackbody continuum suggests an effective photosphere temperature of $6700 \pm 500$ K.}
\end{table}

Using these models, we first fit the continuum contribution by the central blackbody. This yields an effective photosphere temperature of $6700 \pm 500$ K, and this component shown in \autoref{Figure:COFit}a. The fit was carried out using the non-linear least squares fitter \texttt{curve\_fit} from Python’s \texttt{scipy} package \citep{virtanen20}. The resulting fit reveals a clear excess in the continuum past $\sim$$2.2$ $\mu$m, which we attribute to emission from warm dust.

For the second continuum component, we fit a warm dust continuum using the modified blackbody model described by \autoref{Equation:DustModel}. We assume carbonaceous dust \citep{mutschke04} with a grain size of 0.01 $\mu$m. We restrict the model to fit the continuum in narrow regions ($2.172 - 2.183$ $\mu$m and $2.225 - 2.238$ $\mu$m). This is due to the CO features becoming prominent beyond 2.25 $\mu$m and extending past the end of our spectral coverage ($\sim$$2.4$ $\mu$m) and to avoid contamination from the Na I and He I lines. To fit this modified blackbody component, we used the non-linear least-squares fitting routine \texttt{kmpfit} from the \texttt{Kapteyn} package \citep{terlouw14}.

\begin{table*}
    \begin{center}
    \caption{CO Detections in SESNe}
    \label{Table:CODetections}
    \vspace{-0.4cm}
        \begin{tabular}{ccccccccc}
        \hline \hline
        SN                & Type          & $t_{\text{First CO}}$$^a$ & $t_{\text{Last Non-Obs.}}$ & T$_{CO}$    & FWHM$_{CO}$ & M$_{CO}$          & T$_{d}$$^b$ & M$_{d}$              \\
                          &               & (days)                    & (days)                     & (K)         & (\kms)      & (10$^{-4}$ \msun) & (K)    & ($10^{-5}$ \msun)         \\
        \hline
        $\sim$15 SNe      & IIP           & $>$100                    & $>$53                                                                                                           \\
        1987A             & II-pec        & $129-600$                 & $>$100                     & $3200-1200$ & $2000-1200$ & $1-100$   & $10,000-25$$^c$ & $10 - 70000\,^c$         \\
        2017eaw           & IIP           & $124-205$                 & $>$100                     & $3000-2700$ & $2800-2750$ & $0.6-2$   & $1400-1200$     & $0.1 - 10$               \\
        2023ixf           & II            & $199-307$                 & $>$71                      & $2300-1900$ & $3300-3000$ & $0.3-2$   & $\sim$950       & $>$1                     \\
        2024ggi           & II            & $285-385$                 & ...                        & $2400-1900$ & $3200-3000$ & $87-13$   & ...             & ...                      \\
        \hline
        2011dh            & IIb           & 89                        & ...                        & ...         & ...         & ...       & ...             & ...                      \\
        2024uwq           & IIb           & 76                        & ...                        & ...         & ...         & ...       & ...             & ...                      \\
        \hline
        2000ew            & Ic            & 97                        & 39                         & 2000        & 2500        & ...       & ...             & ...                      \\
        2007gr            & Ic            & 82                        & 58                         & ...         & ...         & ...       & ...             & ...                      \\
        2013ge            & Ic            & 64                        & 56                         & ...         & ...         & ...       & ...             & ...                      \\
        2016adj           & Ic$^d$        & 74                        & 37                         & 5100        & 3500        & 1.7       & ...             & ...                      \\
        2020oi            & Ic            & 63                        & 29                         & 3150        & 3700        & 0.8       & 810             & 10                       \\
        2021krf           & Ic            & 68                        & 43                         & ...         & ...         & ...       & 1000            & 2                        \\
        \textbf{2024rbc}  & \textbf{Ib}   & \textbf{62}               & \textbf{47}                & 4040 (435)  & 5905 (1960) & 5.2 (1.2) & 910 (10)        & 130 (10)                 \\
        \hline
        2022wnt           & SLSN I        & 310                       & 150                        & ...         & ...         & ...       & 1000            & 10                       \\
        \hline \hline 
        \end{tabular}
   \end{center}
\renewcommand{\baselinestretch}{0.8}
\vspace{-0.25cm}
\footnotesize{$^a$Days are measured relative to the explosion date of each SN. For SESNe, the reported value corresponds to the first CO detection, whereas for Type II SNe, the values (including CO and dust parameters) indicate the range between the first and last CO detections, listed in chronological order. References are: Type II SNe \citep[][references therein]{banerjee18, sarangi18, rho18sn}, including 1987A \citep{liu92, cherchneff09, wooden93, wesson21}, 2017eaw \citep{rho18sn,tinyanont19},2023ixf \citep{park25, medler25}, 2024ggi \citep{dessart25, mera26}, 2011dh \citep{ergon15}, 2024uwq \citep{subrayan25}, 2000ew \citep{gerardy02}, 2007gr \citep{hunter09}, 2013ge \citep{drout16}, 2016adj \cite{banerjee18}, 2020oi \citep[][$M_{CO} > $10$^{-3}$ \msun\ from non-LTE model]{rho21}, 2021krf \citep{ravi23}, 2022wnt \citep{tinyanont23}.}\\
\footnotesize{$^b$ Most dust mass estimates are derived from NIR spectroscopy at early phases ($\lesssim 600$ days), with the notable exception of SN~1987A \citep[see][and references therein]{wesson21}.}\\
\footnotesize{$^c$ The dust parameters of SN\,1987A span from early (+260 d) to late (+23 yr) phases \citep{wooden93,matsuura11,sarangi18}.}\\
\footnotesize{$^d$SN 2016adj was reclassified as Type Ic \citep{stritzinger24, stritzinger23} from Type IIb \citep{holoien17, banerjee18}. The CO detection phase determined by \cite{stritzinger24} precedes the estimate of \cite{banerjee18} by 16 days.}
\end{table*}

The resulting dust continuum fit with all three components is shown in \autoref{Figure:COFit}b. From this fit, we derive a CO temperature of $4040 \pm 435$ K and a mass of $(5.2 \pm 1.2) \times 10^{-4}$ $M_{\odot}$. The CO \new{has a high velocity width} of $5905 \pm 1960$ \kms, agreeing with the absence of clearly distinguishable band heads. \new{Fitting this model also yields a dust mass of $(1.3 \pm 0.1) \times 10^{-3}$ $M_{\odot}$ and a dust temperature of $910 \pm 10$ K. The temperature is within the expected temperature range of warm dust ($700 - 1100$ K).} Overall, this multi-component model provides a good match to the observed spectrum.


Furthermore, to test the robustness of our CO parameters against the assumption of a warm dust continuum, we also fitted for CO emission resting on a constant continuum. The constant continuum was fit using the \texttt{kmpfit} module over the same wavelength ranges as the warm dust continuum. \autoref{Figure:COFit}c displays this simplified model, which derives a CO temperature of $4170 \pm 390$ K and a mass of $(6.1 \pm 1.2) \times 10^{-4}$ $M_{\odot}$. This model indicates \new{an even higher CO velocity width}: $7970 \pm 1620$ \kms. Like the previous model, the observed spectrum is closely fit. The derived parameters of both CO fits are listed in \autoref{Table:COFitParametersNew}.

Taking both continuum estimates into consideration, the CO should be at $\sim$$4000$ K, between $6000$ and $8000$ km s$^{-1}$, and have a mass of $\sim$$5.5\times10^{-4}M_{\odot}$. There is reasonable agreement between the derived CO values of the two models, indicating that they are largely independent of our continuum estimations. 

\subsection{CO Emission in Supernovae}

\new{\autoref{Table:CODetections} summarizes known CO detections in SNe. More than a dozen Type II/IIP SNe have shown CO emission, including SNe 1987A \& 2017eaw, and more recent events such as SNe 2023ixf \citep{park25, medler25} and 2024ggi \citep{dessart25, mera26}.}

\new{In Type II SNe, CO is typically detected $\sim$100 days post-explosion, with a detection window generally spanning $100-600$ days. The onset is remarkably consistent with the predictions of chemically controlled models of molecule and dust formation \citep{sarangi13, sarangi15, sluder18}. CO masses derived using LTE models are in the range of $1\times10^{-4}$ to $5\times10^{-3}$ $M_\odot$ \citep[][and references therein]{rho18sn}, while non-LTE models yield higher values, ranging from $1\times10^{-3}$ to $5\times10^{-2}\ M_\odot$. Recent estimates for SN\,2024ggi give $8.7\times10^{-3}$ and $1.3\times10^{-3}$ $M_\odot$ at days 285 and 385, respectively.}

\new{Non-LTE CO modeling combines explosion models with the time evolution of molecular formation \citep{hoeflich88, liu92} for specific SN types. More sophisticated approaches incorporate multidimensional treatments of molecule-forming regions \citep{liljegren20}. In contrast, LTE models are computationally simpler and less physically complete; however, they can be applied across different SN types, enabling direct comparison of results \citep[][and references therein]{goorvitch94, rho18sn} like in Table \ref{Table:CODetections}.}


The detection of CO overtone emission in SN\,2024rbc
is of particular interest as it is the first observation of such a feature in a Type Ib SN. \autoref{Table:CODetections} lists a number of CO detections in other SESNe for comparison, where at least six Type Ic SNe have detected CO. Additionally, \rbc\ is the earliest confirmed observation in \autoref{Table:CODetections}. 
\new{SESNe exhibit higher CO temperatures and broader line widths than Type II SNe, possibly reflecting earlier CO formation. The CO temperature and velocity width of \rbc\ is comparable to those of the two other SESNe, 2016adj and 2020oi.}
The CO mass derived assuming non-LTE is expected to exceed the LTE estimate by more than an order of magnitude \citep{liu92, cherchneff08}. For example, in the case of SN\,1987A, the non-LTE CO mass was found to be $40-100$ times larger than the corresponding LTE value \citep{liu92}. By analogy, we therefore expect the CO mass of \rbc\ to be on the order of $\sim$$10^{-3}$ $M_{\odot}$.

Why is it rare to detect CO in a Type Ib SN compared to Type Ic? Examining the population of Type Ib and Ic SNe in the Transient Name Server (TNS) during 2024--2025 with discovery magnitudes $<21$ mag, we find that the numbers are comparable. However, when we restrict to $<19$ mag, the number of Type Ib SNe is about 20\% smaller than Type Ic. Thus, it is possible that Type Ib SNe have a slightly smaller population, but note that the TNS may not contain a complete sample. Furthermore, Type Ib SNe show strong helium emission, although most of the detected lines are from neutral helium. It is possible that some helium is ionized (He$^{+}$), which affects the timing of molecular formation. \cite{cherchneff25} showed that when sufficient He$^{+}$ is present, CO formation is suppressed.

Now that early CO formation been confirmed in the Type Ib \rbc, the need for further observational study to quantify its impact on efficient dust formation is strongly emphasized. 
\new{Theoretical models are essential for interpreting the differences among Type Ib, Ic, and other SNe regarding CO formation, cooling processes, and the connection to dust formation.}
Observing the first CO overtone implies strong emission at the CO fundamental and consequently significant CO formation/excitation. The presence of CO at early times, given its function as a coolant, lends credence to \rbc's potential for productive dust formation.


\subsection{Dust Emission in Supernovae}

\new{\autoref{Table:CODetections} also presents the dust parameters of SESNe in comparison with those of several well-studied Type II SNe. In \rbc, we found a dust temperature of 910 K and dust mass of $1.3\times10^{-3}$ $M_\odot$. This dust mass is more than an order of magnitude larger than those found in SESNe 2020oi \citep{rho21} and 2021krf \citep{ravi23} based on NIR spectroscopy (up to 2.5 $\mu$m).}

\new{The dust continuum has been examined to explore several possible origins, including freshly formed dust in the ejecta, heated pre-existing dust in the circumstellar medium, and infrared echoes \citep[][and references therein]{rho18sn, rho21, tinyanont19}.
The presence of CO is a direct evidence for onset of dust formation in the ejecta of a SN because CO is a dominant coolant, thus paving the way for dust grains with lower nucleation temperatures (e.g. carbonaceous grains) to form. The dust mass measured in Type Ibn SN\,2006jc at day 200 was roughly $0.5 - 2 \times 10^{-3}$ $M_\odot$ \citep{nozawa08}, which is comparable to \rbc, but at a later time.}

The dust mass estimated by fitting ($\sim$$10^{-3}$ $M_{\odot}$) by itself is certainly insufficient to answer to the task of dust generation in the early Universe, which would require (at minimum) a total formed dust mass $>$0.1 $M_{\odot}$ \citep{nozawa03, todini01, sluder18}. However, as these measurements of SNe 2024rbc and 2006jc are conducted on a very early spectrum, there is significant room for growth. 

\new{The dust formation model for Type Ib SNe predicts a dust mass of $\sim$1 $M_\odot$, as the ejecta in Type Ib SNe cool more rapidly than in Type II SNe. This is consistent with the earlier CO formation timescales observed in SESNe \citep{nozawa08}.}
Figure 5 in \citealt{tinyanont25} provides a helpful guide to the potential of \rbc\ in terms of dust formation. Type Ib SN\,2014C initially registered a dust mass of $\sim$$10^{-3}$ $M_{\odot}$ at approximately 250 days, which is comparable to the dust found in \rbc\ at only 62 days. Considering that SN\,2014C was measured to have a dust mass of $\sim$$10^{-1}$ $M_{\odot}$ after $\sim$3500 days, we argue that \rbc\ shows potential for significant dust formation over the next decade.
This possibility invites follow-up observations of \rbc\ and other, new Type Ib SNe on this timescale.

Further examination of dust formation of SESNe, including \rbc, would require observations by a spectrograph with mid--infrared (MIR) capabilities, \new{such as LRS or MRS on JWST's MIRI.} The dust mass estimates using NIR observations alone should be considered the lower limit. \new{As seen in the JWST observations of the Type II SN\,2023ixf, significant MIR dust emission has been detected alongside molecular CO emission \citep{medler25}. The inferred dust mass may be comparable to that of SN\,1987A ($0.1 - 0.7\ M_\odot$), as listed in Table~\ref{Table:CODetections}.}

\new{It is unfortunate that only a limited number of JWST observations of SESNe, including mid-infrared (MIR) spectra, have been obtained and published in literature. The contribution of supernovae to dust production in the early Universe from high-mass ($\sim$$40$ $M_\odot$) stars is predicted to be comparable to, or even higher than, that from lower-mass stars (8–15 $M_\odot$, such as Type IIP progenitors) after the initial mass function, nucleosynthesis yields, and dust production rates are taken into account \citep{rho23}.}

\new{Therefore, JWST near- and mid-infrared observations, together with ground-based near-infrared observations of a larger sample of SESNe (including Type Ib, Ic-BL, and Ibn/Icn events), and improved constraints on CO and dust formation in these systems, are crucial for determining the dust composition, accurately estimating the total dust mass from both the warm and cool dust components, and ultimately addressing the fundamental question of whether supernovae were the primary producers of dust in the early Universe.}



\section{Conclusions}\label{Section: Conclusions}
Our conclusions are as follows:
\begin{enumerate}
  \item The light curves of \rbc\ show clear, singular peaks at $12-18$ days post-explosion and a two phase decline afterwards. The $r$ band light curve of \rbc\ is in strong agreement with other Type Ib and some Type Ic SNe.
  
  \item Fitting model light curves from a He star progenitor indicate a pre-explosion He star mass of $3.1$ $M_{_\odot}$ with M($^{56}$Ni)$=0.07$ $M_{\odot}$, E$_{\text{Explosion}}=10^{51}$ erg, $\text{M}_{\text{Ejecta}}=1.7$ $M_{\odot}$, and the $^{56}$Ni-mixing parameter $f_m=0.5$. Furthermore, the absence of bright optical emission at very early times suggests that the CSM around \rbc\ was moderate at the time of the explosion.
  
  \item The spectra of \rbc\ exhibit numerous atomic and ionized metal lines (Mg I, Ca I, Si I, Ca II, Si II, and Fe II) and several uncommon lines (C I, Na I). Strong and broad Ca II and Mg I emission, absorption, and P-Cygni profiles are present in the optical and NIR spectra. Comparison of the spectral evolution of \rbc\ to other Type Ib and Ic SNe support the classification of \rbc\ as a Type Ib SN.
  
  \item At 62 days post-explosion, a K-band continuum emission indicative of warm dust is evident in the NIR spectra. Additionally, the first CO overtone appears at this time and is the first confirmed observation of CO in a Type Ib SN. LTE modeling of the CO overtone gives a temperature of $\sim$$4000$ K, \new{a velocity width} between $6000$ and $8000$ km s$^{-1}$, and a mass between $5$ and $6\times10^{-4}$ $M_{\odot}$. Modeling the dust continuum indicates a dust temperature of $910 \pm 10$ K and a mass of $(1.3 \pm 0.1) \times 10^{-3}$ $M_{\odot}$.
  
  \item The early detection of CO and dust in the Type Ib \rbc\ lends further credence to the idea that rapid, efficient dust formation is possible in SESNe. Follow-up observations and further study of this class of SNe are necessary to better understand the relationship between CO and dust formation and the plausibility of dust formation via CCSNe in the early Universe. 
\end{enumerate}

We thank Yuxin Dong for participating in the Keck observations of the optical spectrum.
Part of the data presented was obtained with the international Gemini Observatory, a program of NSF NOIRLab. This program is managed by the Association of Universities for Research in Astronomy (AURA) under a cooperative agreement with the U.S. National Science Foundation on behalf of the Gemini Observatory partnership: the U.S. National Science Foundation (United States), National Research Council (Canada), Agencia Nacional de Investigaci\'{o}n y Desarrollo (Chile), Ministerio de Ciencia, Tecnolog\'{i}a e Innovaci\'{o}n (Argentina), Minist\'{e}rio da Ci\^{e}ncia, Tecnologia, Inova\c{c}\~{o}es e Comunica\c{c}\~{o}es (Brazil), and Korea Astronomy and Space Science Institute (Republic of Korea).
Additionally, a part of the data presented here was obtained with ALFOSC, which is provided by the Instituto de Astrof\'{i}sica de Andaluc\'{i}a (IAA) under a joint agreement with the University of Copenhagen and NOT. 
W. M. Keck Observatory access was supported by Northwestern University and the Center for Interdisciplinary Exploration and Research in Astrophysics (CIERA).

The participation of R.H. was made possible by the SETI Institute REU program (NSF grant \#\,2447895).
J.R. was in part supported by a NASA ADAP grant (80NSSC23K0749) and Brain Pool visiting program for Outstanding Overseas Researchers by the National Research Foundation of Korea (NRF-2022H1D3A2A01096434).
S.-C.Y. and S.H.P. were supported by the NRF RS-2024-00356267.
C.L. was supported by DoE award \#\,DE-SC0025599 to Northwestern University.

\vskip 0.15truecm
\textit{Facilities}: 
ATLAS-HKO,
Gemini (GNIRS),
IRTF (SpeX),
Keck (LRIS, NIRES),
Lick (Kast, Nickel),
NOT (ALFOSC),
\new{Pan-STARRS},
ZTF (SEDM, ZTF-Cam)

\textit{Software}:
Astropy \citep{astropy:2013, astropy:2018, astropy:2022},
NumPy \citep{harris20},
Figaro \citep{shortridge92}, 
Gemini IRAF Package \citep{fitzpatrick25},
IRAF \citep{tody86}, 
Kapteyn \citep{terlouw14},
Matplotlib \citep{hunter07},
PypeIt \citep{pypeit:joss_pub, pypeit:zenodo},
SciPy \citep{virtanen20},
Spextool \citep{2004PASP..116..362C},
STELLA \citep{blinnikov00, blinnikov06},
UCSC Spectral Pipeline \citep{siebert20},
XDGNIRS \citep{mason15}

\begin{figure}[!hb]
\includegraphics[width=8truecm]{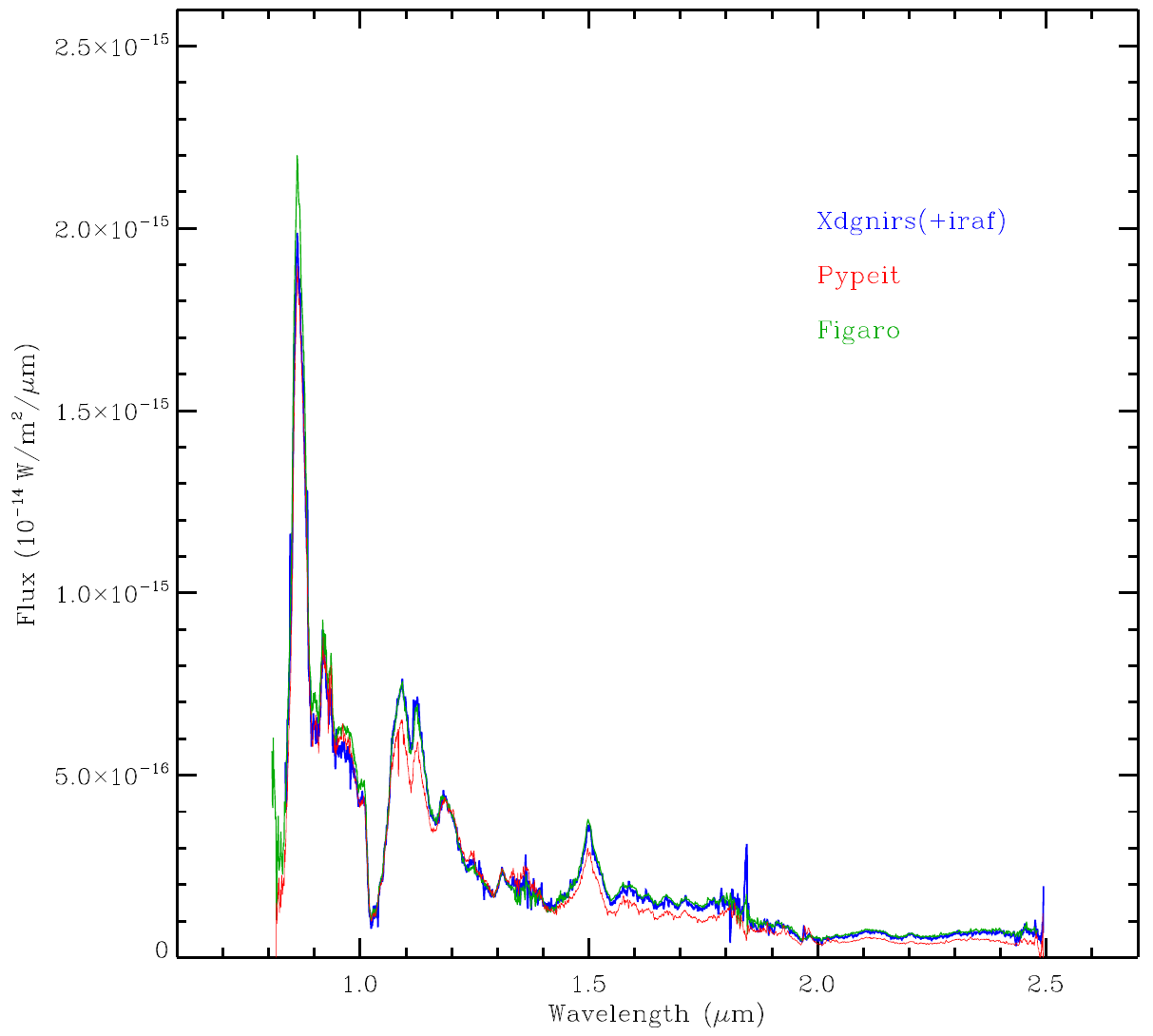}
\caption{A comparison of the reduction of the +62d Gemini GNIRS spectrum by \texttt{xdgnirs} (blue), \texttt{pypeit} (red), and \texttt{Figaro} (green).}
\label{Figure:ReductionComparison}
\end{figure}

\section*{APPENDIX}
\section*{A. Data Reduction Comparisons}\label{Appendix}

We compare the \texttt{xdgnirs}, \texttt{pypeit}, and \texttt{Figaro} reductions of the observation by GNIRS at +62d. The reduction process for \texttt{xdgnirs} was described in Section \ref{Subsection: NIR Spectroscopy}. The process for \texttt{Figaro} was discussed in \citet{rho18sn}. Below, we describe the reduction process for \texttt{pypeit} \citep{pypeit:joss_pub}. 

The raw data was reduced without A–B pairing. The flats and arc frames obtained during the observation were also provided to \texttt{pypeit}. The pipeline performed flat-fielding, wavelength calibration, aperture identification, tracing, and spectral extraction automatically. Flux calibration, order stitching, and telluric correction were then carried out as separate, manually initiated steps within the \texttt{pypeit} toolset following the initial reduction. The nearby A2V star HIP 4129 was used as both a telluric and flux standard to minimize differences in airmass. The final reduced spectrum is continuous over the wavelength range $0.817-2.496$ $\mu$m. Infrared data reduction with \texttt{pypeit} is also described in detail by \cite{tinyanont24}.

The spectra produced by the three reduction methods are shown in \autoref{Figure:ReductionComparison}. We find that they are largely consistent with one another.

\newpage
\bibliography{ms.bib}
\end{document}